\documentclass[prd,twocolumn]{revtex4}
\usepackage{dcolumn}
\usepackage{multirow}
\usepackage{graphicx}
\usepackage{amssymb}
\usepackage{bm}
\usepackage{hyperref}
\usepackage{epstopdf}
\usepackage{color}
\usepackage{mathrsfs}
\usepackage{amsmath,amssymb,amsthm}
\usepackage{rotating}
\usepackage{sverb, longtable}
\usepackage{subfigure}

\usepackage[graphicx]{realboxes}
\usepackage{adjustbox}

\begin{document}

\title{On a robust analysis of the growth of structure}

\author{Deng Wang}
\email{dengwang@ific.uv.es}
\affiliation{Instituto de F\'{i}sica Corpuscular (CSIC-Universitat de Val\`{e}ncia), E-46980 Paterna, Spain}
\author{Olga Mena}
\email{omena@ific.uv.es}
\affiliation{Instituto de F\'{i}sica Corpuscular (CSIC-Universitat de Val\`{e}ncia), E-46980 Paterna, Spain}
\begin{abstract}
Current cosmological tensions show that it is crucial to test the predictions from the canonical $\Lambda$CDM paradigm at different cosmic times. One very appealing test of structure formation in the universe is the growth rate of structure in our universe $f$, usually parameterized via the growth index $\gamma$,  with $f\equiv \Omega_m(a)^\gamma$ and $\gamma \simeq 0.55$ in the standard $\Lambda$CDM case. Recent studies have claimed a suppression of the growth of structure from a variety of cosmological observations, characterized by $\gamma>0.55$. By employing different self-consistent growth parameterizations schemes, we show here that $\gamma<0.55$, obtaining instead \emph{an enhanced growth of structure today}. This preference reaches the $3\sigma$ significance using Cosmic Microwave Background observations, Supernova Ia and Baryon Acoustic Oscillation measurements. The addition of Cosmic Microwave Background lensing data relaxes such a preference to the $2\,\sigma$ level, since a larger lensing effect can always be compensated with a smaller structure growth, or, equivalently, with $\gamma>0.55$. We have also included the lensing amplitude $A_{\rm L}$ as a free parameter in our data analysis, showing that the preference for $A_{\rm L}>1$ still remains, except for some particular parameterizations when lensing observations are included. We also not find any significant preference for a multipole dependence of $A_{\rm L}$. To further reassess the effects of a non-standard growth, we have computed by means of N-body simulations the dark matter density fields, the dark matter halo mass functions and the halo density profiles
for different values of $\gamma$. Future observations from the Square Kilometer Array, reducing by a factor of three the current errors on the $\gamma$ parameter, could finally settle the issue. 
\end{abstract}
\maketitle

\section{Introduction}

Cosmological observations can be optimally  fitted within the minimal $\Lambda$CDM model, described by six fundamental parameters. This model has successfully explained a large number of measurements at different scales. Nevertheless,  there are a number of tensions. The Hubble constant one~\cite{Riess:2022mme}, is the most significant tension ($5\sigma$), and it implies the mismatching between the value of the Hubble constant obtained from Planck-2018 CMB observations ($H_0=67.4\pm0.5$ km~s$^{-1}$~Mpc$^{-1}$~\cite{Planck:2018vyg}) and the value of $H_0$ from local measurements using SNIa calibrated with Cepheid variable stars ($H_0=73\pm1$ km~s$^{-1}$~Mpc$^{-1}$~\cite{Riess:2021jrx}). A second conundrum, even if less significant, is the $S_8$ tension, parameter closely related to the matter clustering of matter in the Universe.  $S_8$ could be inferred from the measurements of CMB anisotropies as those from Planck, and, more directly, from the measurements of galaxy lensing made by large surveys such as the Dark Energy Survey (DES) \cite{DES:2021vln,DES:2022ygi,des}, and the Kilo-Degree Survey (KiDS)\cite{Heymans:2020gsg,KiDS:2020ghu,kids}. While Planck CMB data favors a larger value of $S_8$, galaxy surveys (the DES \cite{DES:2021vln,DES:2022ygi,des} and the KiDS-1000 surveys \cite{Heymans:2020gsg,KiDS:2020ghu,kids}) prefer a lower one, leading to a $2\sim3\,\sigma$ discrepancy. 

The so-called lensing anomaly~\cite{Planck:2018vyg,Motloch:2018pjy} is due to the fact that Planck CMB data shows a \emph{preference} for additional lensing. The parameter $A_{\rm L}$~\cite{Calabrese:2008rt}, the so-called lensing amplitude, is a pure phenomenological parameter that artificially increases ($A_{\rm L}>1$) or decreases ($A_{\rm L}<1$) the amount of CMB lensing. If $\Lambda$CDM consistently describes all CMB data, observations should point to $A_{\rm{L}}=1$. The analysis of CMB temperature and polarization anisotropies suggests $A_\mathrm{L} > 1$ at 3$\sigma$ level. CMB lensing also introduces a non-trivial four-point correlation function, and therefore it can be independently measured. Adding this information diminishes the tension, albeit the value of the lensing amplitude is still above the canonical one at about $2\sigma$.

It is therefore extremely important to further test the $\Lambda$CDM paradigm at different epochs and scales. In this regard, future galaxy and weak lensing surveys will be able to test possible departures from the standard cosmological picture.
Among the most appealing theoretical alternatives are those relying on modifications of General Relativity at ultra large length scales. Usually, departures from General Relativity are parameterized via the growth rate $f$~\cite{Wang:1998gt,Linder:2005in,Linder:2007hg}:
\begin{equation}
f \equiv \frac{d}{dln a}\frac{\delta(k,a)}{\delta(k, a_i)} =\Omega_m(a)^\gamma~, 
\label{eq:f}
\end{equation}

\noindent with
\begin{equation}
\Omega_{m}(a)=\Omega_{m0}a^{-3}/(\frac{H^2}{H_0^2})~,
\end{equation}
and, assuming a flat universe:
\begin{equation}
1=\Omega_{m}(a)+\Omega_{DE}(a)~.
\end{equation}
Within the canonical general relativity scheme, and a flat $\Lambda$CDM background, $\gamma=0.55$~\cite{Linder:2005in}. Consequently, a departure from this standard value would suggest an inconsistency between the
concordance cosmological model and observations.
Recently, the authors of \cite{Nguyen:2023fip} have found a higher gravitational growth index than the canonical  value $\gamma = 0.55$ when combining Planck CMB data with weak lensing, galaxy clustering, and cosmic velocity observations. All in all, they find $\gamma = 0.633 ^{+0.025}_{-0.024}$, rejecting therefore with a significance of $3.7\sigma$ the standard value of $\gamma$. The analysis of Ref.~\cite{Nguyen:2023fip} makes use of a simple parameterization in terms of a unique parameter $\gamma$  modifying the power spectrum as follows:

\begin{equation}
    P (\gamma, k, a) = P (k, a = 1) D^2(\gamma, a)~,
\end{equation}

\noindent where the linear growth factor is related to the growth rate, Eq.~(\ref{eq:f}), as
\begin{equation}
    D(\gamma, a) = \exp -\left(\int_a^1 da \frac{\Omega_m(a)^\gamma}{a}\right)~.
\end{equation}

Here we follow a different avenue and revise the current constraints on the growth of structure by means of an alternative parameterization and also considering CMB constraints both alone and in combination with other cosmological observations. Therefore, in order to include possible deviations from general relativity, instead of modifying the matter power spectrum via the growth factor $f$, we make use of a self-consistent parameterization in terms of two functions, $\mu(k,a)$ and $\eta(k,a)$, which modify the Poisson equation and introduce a gravitational slip, respectively:
\begin{eqnarray}
\label{eq:Poisson} 
-k^2 \Phi(k,a) &\equiv& 4\pi G a^2 \mu (k,a) \rho (k,a) \delta (k,a)~;\\
\eta(k,a)     &\equiv& \Psi(k,a)/\Phi(k,a), \label{eq:aniso}
\end{eqnarray}
where $\rho(a)$ is the average dark matter density, $\delta(k,a)$ is  the comoving matter density
contrast and $\Phi$ and $\Psi$  are  the gauge-invariant Bardeen potentials in the
Newtonian gauge:
\begin{equation}
    ds^2 = -(1+2 \Phi) dt^2 + a^2 (1-2\Psi) dx^2~.
\end{equation}

Using effective quantities like $\mu$  and $\eta$ has the advantage that they are able to model any deviations of the perturbation behaviour from $\Lambda$CDM expectations, they are relatively close to observations, and they can also be related to other commonly used parameterizations. 
Indeed, there is a direct relation between $\mu$ and $\gamma$~\cite{Pogosian:2010tj}, in such a way that it is always possible to map one parameterization ($\mu$,$\eta$) into another related one ($\gamma$,$\eta$):

\begin{equation}
\mu = \frac{2}{3} \Omega_m (a)^{\gamma-1}
\left[\Omega_m(a)^\gamma +2 -3\gamma + \frac{3}{2} (\gamma-1) \Omega_m(a)\right]~,
\label{eq:gamma}
\end{equation}
for a flat $\Lambda$CDM universe and constant $\gamma$.
In the following, we shall follow a  parameterization in which the time evolution is related to the dark energy density fraction, i.e. a ‘late-time’ parameterization~\cite{Casas:2017eob}, which also neglects the possible scale dependence of the functions $\mu$ and $\eta$:
\begin{eqnarray}
\label{eq:mueta}
\mu(a)&=&1+E_{11}\Omega_{DE}(a)~;\nonumber \\
\eta(a)&=&1+E_{22}\Omega_{DE}(a)~.
\label{eq:eta}
\end{eqnarray}
If $\mu_0=\mu(a=1$) and $\eta_0=\eta(a=1)$, $(E_{11}, E_{22})$ is equivalent to $(\mu, \eta)$ at $a=1$ today. We shall show our results for this parameterization. We also show results for the case in which we use ($\gamma$, $\eta$)  as parameters describing the deviations with respect to the standard $\Lambda$CDM cosmology, by means of Eqs.~(\ref{eq:gamma}) and (\ref{eq:eta}) as well as for the simplest one parameter $\gamma$ case. 
The structure of the manuscript is as follows. Section~\ref{sec:metholody} describes the numerical codes employed to develop our analyses and also the cosmological observations used along this study.
Section~\ref{sec:results} presents the constraints within the different growth of structure parameterizations from current data, the results from N-body simulations, and the future prospects  for a Square Kilometer Array (SKA)-like survey. Finally, we draw our conclusions in Sec.~\ref{sec:conclusions}. 

\section{Methodology}
\label{sec:metholody}
In order to study the constraints achievable by current CMB and large scale structure probes, we make use of the publicly available code \texttt{MGCAMB}~\cite{Zhao:2008bn,Hojjati:2011ix,Zucca:2019xhg,Wang:2023tjj}, which is a modified version of \texttt{CAMB} \cite{Lewis:1999bs} for cosmic structure growth, and incorporate the $(\gamma,\,\eta)$ growth parameterization into it.  We employ the Monte Carlo Markov Chain (MCMC) method to infer the posterior distributions of models parameters by using the code \texttt{CosmoMC}~\cite{Lewis:2002ah}. We analyze the MCMC chains via the public package \texttt{Getdist} \cite{Lewis:2019xzd}. Notice that we adopt the potential reduction scale factor $R-1=0.03$ proposed by Gelman and Rubin~\cite{Gelman:1992zz} as the convergence criterion of our MCMC analysis. For all the growth parameterizations considered in this analysis, we choose the following uniform prior ranges for the different parameters: the baryon fraction $\Omega_bh^2 \in [0.005, 0.1]$, the cold dark matter fraction CDM fraction $\Omega_ch^2 \in [0.001, 0.99]$, the acoustic angular scale at recombination $100\theta_{MC} \in [0.5, 10]$, the amplitude of primordial scalar power spectrum $\mathrm{ln}(10^{10}A_s) \in [2, 4]$, the scalar spectral index $n_s \in [0.8, 1.2]$, the reionization optical depth $\tau \in [0.01, 0.8]$, the CMB lensing amplitude $A_L \in [0, 2.5]$, the growth index $\gamma \in [0, 1]$, the effective gravitational strength $\mu_0 \in [-3, 3]$ and the effective anisotropic stress $\eta_0 \in [-3, 3]$.

Furthermore, to investigate the impacts of the  growth index $\gamma$ on cosmic structure formation, we modify the online software \texttt{Gadget2}~\cite{Springel:2005mi}. Specifically, we modify the Poisson equation by replacing the factor $\mu(k,a)$ in Eq.(\ref{eq:Poisson}) by Eq.(\ref{eq:gamma}), which characterizes the effective gravitational strength in this model. Note that we do not include the effect of anisotropic stress from photons and neutrino species on the large scale structure in this analysis. The background evolution of this model is equivalent to that of $\Lambda$CDM, since this model only considers the structure growth. As a consequence, we just need to consider the scalar density perturbation when making the initial condition. To simulate such a modified universe, at first, we use the best fit values of the model parameters from our conservative cosmological constraints as fiducial parameters. Then, using the code \texttt{2LPTic} \cite{Crocce:2006ve}, we generate the initial condition at redshift $z=49$  with $256^3$ particles and a box size of 200 $h\,\mathrm{Mpc}^{-1}$, and evolve the universe to $z=0$. The softening length we use is 15 $h\,\mathrm{kpc^{-1}}$ in simulations. In order to study the effects of $\gamma$ more clearly on structure formation, we also make a comparison group of simulations by only varying $\gamma$ and fixing other parameters. We use the \texttt{AHF} code \cite{Knollmann:2009pb} to identify the dark halos and generate the halo catalogues in all the simulations. 

Concerning the cosmological and astrophysical observations, our baseline data sets and likelihoods include:

\begin{itemize}

    \item CMB. Observations from the Planck satellite have very important meanings for cosmology and astrophysics. They have measured the matter components, the topology and the large scale structure of the universe. We therefore consider here Planck 2018 temperature and polarization (TT TE EE) data, and the low-$\ell$ temperature and polarization likelihoods at $\ell < 30$, namely TTTEEE+low${\ell}$+lowE~\cite{Planck:2018vyg,Planck:2019nip,Planck:2018nkj,planck}. We refer to this combination as ``C''.

    \item Lensing. The CMB photons traverse almost the entire observable Universe to arrive here today and are deflected by gradients in the gravitational potentials associated with inhomogeneities in the universe. We use the Planck 2018 lensing likelihood~\cite{Planck:2018lbu}, reconstructed from measurements of the power spectrum of the lensing potential. We refer to this dataset as ``L''.
	
    \item BAO. Baryon Acoustic Oscillations (BAO) are very clean observations to explore the evolution of the universe, which are unaffected by uncertainties in the nonlinear evolution of the matter density field and by other systematic uncertainties which may affect other observations. Measuring the positions of these oscillations in the matter power spectrum at different redshifts can place strong constraints on the cosmic expansion history. We employ BAO measurements extracted from the 6dFGS~\cite{Beutler:2011hx}, SDSS MGS~\cite{Ross:2014qpa}, BOSS DR12~\cite{Alam:2016hwk} and eBOSS DR16~\cite{eBOSS:2020yzd,Ross:2020lqz} samples. We refer to this dataset combination as ``B''.
    
    \item SNIa. Supernovae Ia luminosity distances are powerful distance indicators to probe the expansion history of the universe, especially, the equation of state of the dark energy component. We adopt SNIa data points from the largest Pantheon+ sample~\cite{Scolnic:2021amr}, which is made of 1701 light curves of 1550 spectroscopically confirmed SNe Ia coming from 18 different surveys. This updated sample has a significant improvement relative to Pantheon \cite{Pan-STARRS1:2017jku}, specially at low redshifts, and covers the redshift range $z\in[0.000122, 2.26137]$. We refer to this dataset combination as ``S''.
    
\end{itemize}
In what follows, we shall report results for CMB alone (``C''),  CMB plus CMB lensing (``CL''), CMB plus BAO plus Supernovae Ia datasets (``CBS'') and CMB plus BAO plus Supernovae Ia plus CMB lensing (``CBSL'').

Notice that we do not consider weak lensing data. Despite these measurements have been considered in previous works~\cite{Nguyen:2023fip}, in order to properly compute the weak lensing observables one would need to modify accordingly to the parameterization used here (see Eqs.~(\ref{eq:gamma}) and (\ref{eq:mueta})) the halofit~\cite{Smith:2002dz,Bird:2011rb,Takahashi:2012em} numerical code. In the absence of a publicly available modified version of the former code, we follow here a very conservative approach, neglecting the weak lensing input.

The most conservative data combination is therefore what we refer to \textbf{CBS}, i.e. CMB, BAO and SNIa observations. Such a data combination will be enlarged with weak lensing and RSD measurements for the sake of comparison with previous studies~\cite{Nguyen:2023fip}.

To investigate the observational viability of different growth parameterizations, taking $\Lambda$CDM as the reference model, we compute the Bayesian evidences of nine models, $\varepsilon_i$, and Bayes factor, $B_{ij}=\varepsilon_i/\varepsilon_j$, where $\varepsilon_j$ is the statistical evidence of the reference model. According to Ref.~\cite{Trotta:2005ar}, we employ a revised and more conservative version of the so-called Jeffreys' scale, i.e., $\ln B_{ij} = 0 - 1$, $\ln B_{ij} = 1 - 2.5$, $\ln B_{ij} = 2.5 - 5$ and $\ln B_{ij} > 5$ indicate an \textit{inconclusive}, \textit{weak}, \textit{moderate} and \textit{strong} preference of the model $i$ relative to the reference model $j$. It is noteworthy that if $\ln B_{ij} < 0$ for an experiment, it implies that data prefers the reference model.

\begin{table}[!t]
	\renewcommand\arraystretch{1.5}
	\caption{Bayesian factor for different models in light of Planck-2018 CMB data.}
        \setlength{\tabcolsep}{6mm}
	{\begin{tabular}{@{}cccccc@{}} \toprule
			Models            &$\ln B_{ij}$           \\ \colrule
			$\Lambda$CDM       &0                 \\  
            $\Lambda$CDM+$A_L$ &-1.154         \\
            $\Lambda$CDM+$A_\ell$+$A_m$ &-2.511         \\
            $\gamma$           &-0.997            \\
            $\gamma$+$A_L$           &-3.104            \\
            $\gamma,\eta$      &-0.053          \\
            $\gamma,\eta$+$A_L$      &-2.133          \\
            $\mu,\eta$     &1.397          \\
            $\mu,\eta$+$A_L$      &-0.501          \\
            $\mu,\eta$+$A_\ell$+$A_m$      &-2.444          \\
			\botrule
		\end{tabular}
		\label{t1}}
\end{table}

\begin{figure*}
\begin{tabular}{c c}
 \includegraphics[width = 0.5\textwidth]
{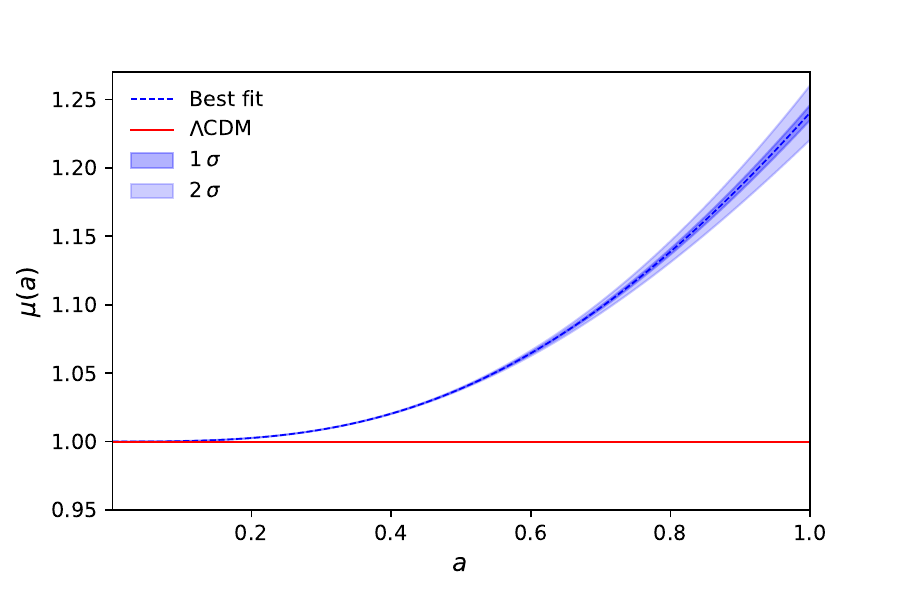} & 
 \includegraphics[width = 0.5\textwidth]
{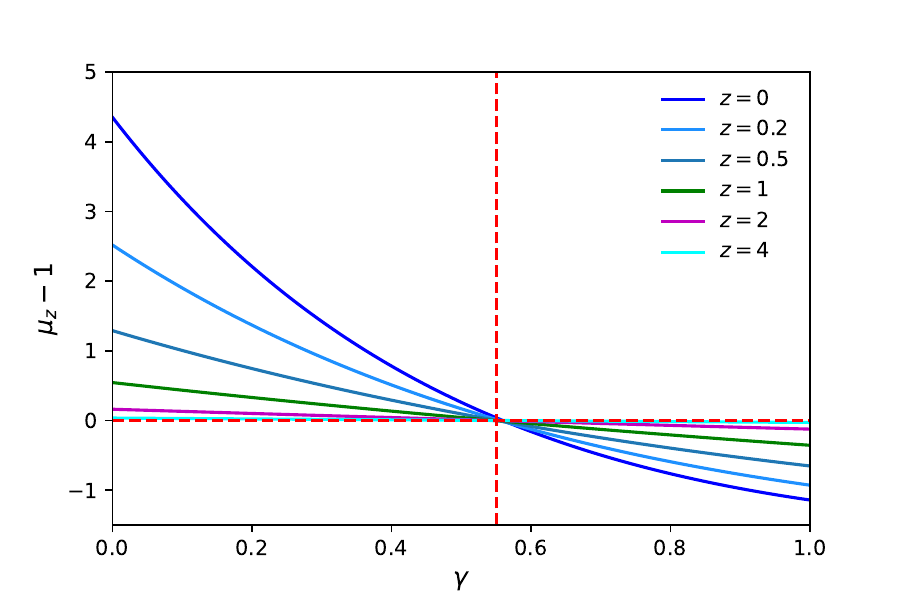} 
\end{tabular}
\caption{\textit{Left panel:} Reconstructed values of $\mu(a)$ together with the 1 and 2$\sigma$ errors from CMB temperature and polarization plus CMB lensing measurements. \textit{Right panel:} Equivalence between the $\gamma$ parameter and $\mu(z)-1$, see Eq.~(\ref{eq:gamma}), for a number of different redshifts. The cross point corresponds to $\Lambda$CDM and the horizontal and vertical dashed lines depict the expectations within standard cosmology, i.e. $\mu=1$ and $\gamma=0.55$ respectively.}
\label{fig:mua}
\end{figure*}

\section{Results}
\label{sec:results}
\subsubsection{Constraints from CMB, large scale structure and Supernovae Ia data}

\begin{figure*}
\begin{tabular}{c c}
 \includegraphics[width = 0.5\textwidth]
{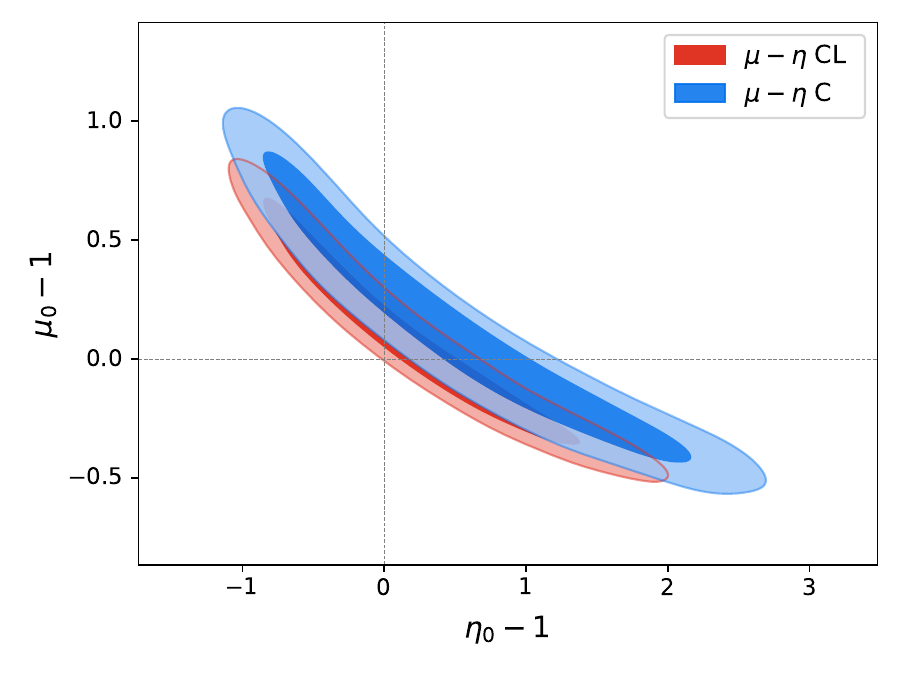} & 
 \includegraphics[width = 0.5\textwidth]
{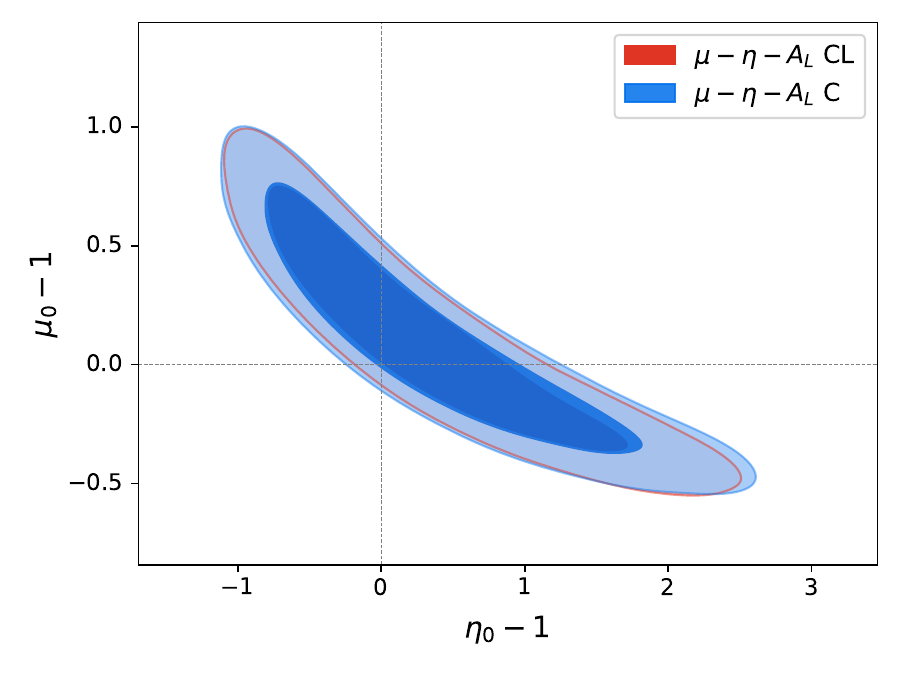} 
\end{tabular}
\caption{\textit{Left panel:} $68\%$ and $95\%$ CL contours in the ($\eta_0-1$, $\mu_0-1$) plane within the growth model parameterized via $\mu$ and $\eta$, see Eqs.~(\ref{eq:mueta}), obtained by using CMB temperature and polarization measurements (blue contours) and CMB temperature,  polarization and lensing observations (red contours). \textit{Right panel:} As in the left panel but when the lensing amplitude $A_L$ is also considered as a freely varying parameter.}
\label{fig:mueta}
\end{figure*}

\begin{figure*}
\begin{tabular}{c c}
 \includegraphics[width = 0.5\textwidth]
{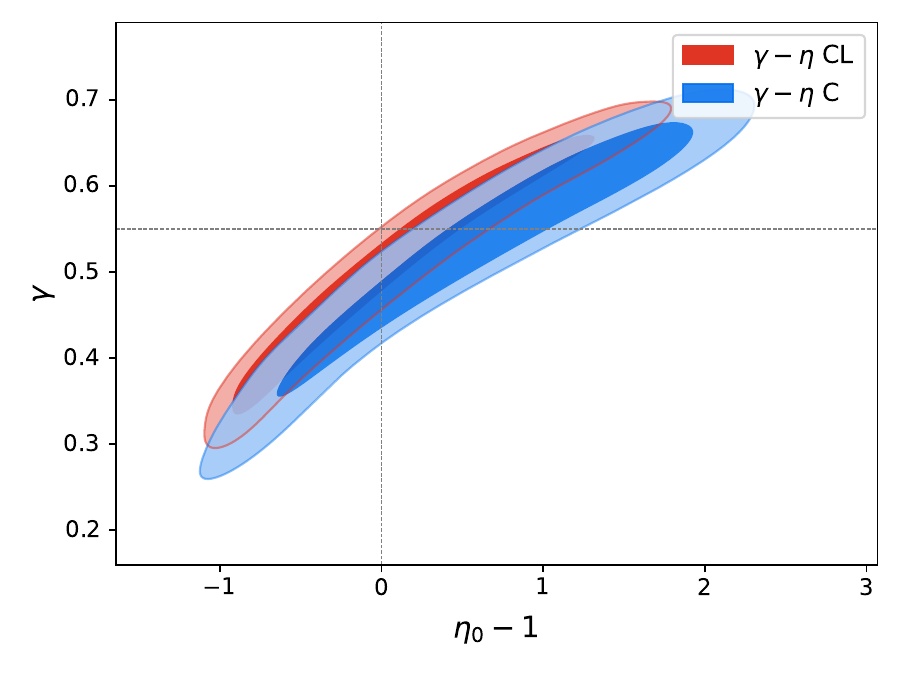} & 
 \includegraphics[width = 0.5\textwidth]
{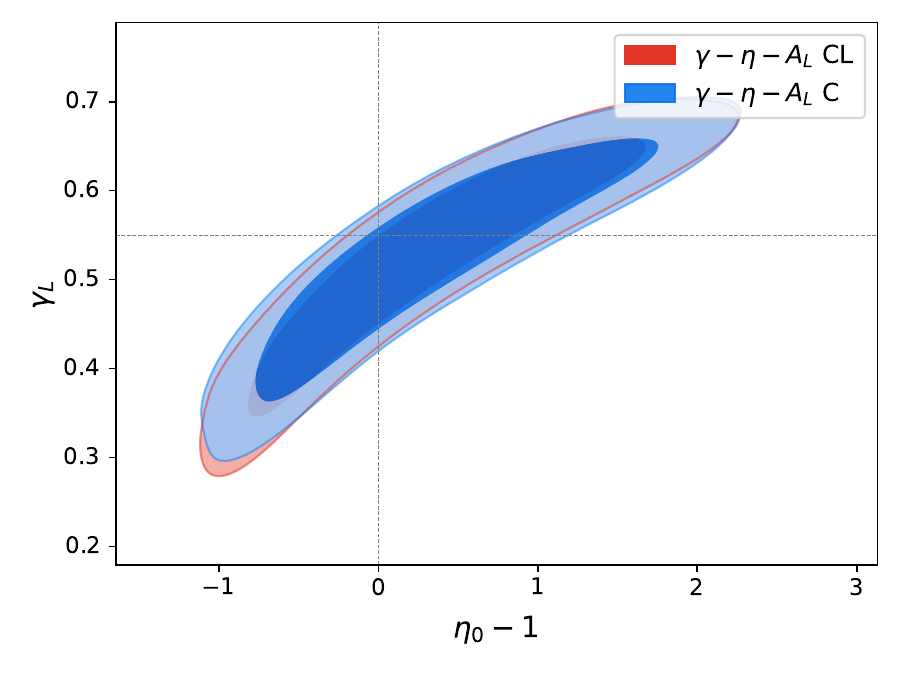} 
\end{tabular}
\caption{\textit{Left panel:} $68\%$ and $95\%$ CL contours in the ($\eta_0-1$, $\gamma$) plane within the growth model parameterized via via $\gamma$ and $\eta$, see Eq.~(\ref{eq:gamma}) and the second of Eqs.~(\ref{eq:mueta}) obtained by using CMB temperature and polarization measurements (blue contours) and CMB temperature,  polarization and lensing observations (red contours). \textit{Right panel:} As in the left panel but when the lensing amplitude $A_L$ is also considered as a freely varying parameter.}
\label{fig:gammaeta}
\end{figure*}

Table \ref{t1} summarizes our results concerning the Bayesian preference for different models versus the canonical $\Lambda$CDM scenario arising from CMB observations, including also CMB lensing in our analyses. Notice that a model with both ($\mu$, $\eta)$  as free parameters is substantially-strongly preferred over the canonical scenario. We show the reconstructed $\mu(a)$ in the left panel of Fig.~\ref{fig:mua} from CMB plus CMB lensing measurements. The right panel depicts the equivalence between the $\gamma$ parameter and the $\mu(z)$ one: an enhancement of the growth of structure today, characterized by $\gamma<0.55$, implies $\mu(z)>1$ and $\mu_0-1>0$. As it is clear from the left panel of Fig.~\ref{fig:mua}, since a value of $\mu>1$ is  reconstructed from CMB temperature, polarization and lensing data, this would imply an enhancement of the growth of structure. Concerning the parameterization in terms of ($\mu$, $\eta)$, it is almost equally favoured than the $\Lambda$CDM cosmology. Notice however that the parameterization which only makes use of the $\gamma$ parameter is disfavoured. The left panels of Figs.~\ref{fig:mueta} and \ref{fig:gammaeta} depict the 68\% and 95\%~CL allowed contours from  Planck CMB data, with and without CMB lensing, in the ($\mu_0-1$, $\eta_0$-1) and ($\gamma$, $\eta_0$-1) planes, associated to the  ($\mu$, $\eta$) and ($\gamma$, $\eta$) parameterizations respectively.
We start reporting the values of the most relevant parameters in what follows. Table~\ref{tab:gamma} depicts the results for the simplest growth parameterization in terms of a single parameter $\gamma$, for the sake of comparison with previous results\cite{Nguyen:2023fip}. Notice that, contrarily to the results previously quoted in the literature,  we obtain always $\gamma<0.55$, i.e. \emph{an enhanced growth of structure today}, in agreement also with the results depicted in Figs.~\ref{fig:mueta} and \ref{fig:gammaeta}. Notice that CMB alone prefers $\gamma<0.55$ at $3\sigma$. The addition of SNIa and BAO measurements do not change this preference, while the addition of CMB lensing relaxes the preference to the $2\sigma$ level. This is related to the fact that the lensing data restores the value of the lensing amplitude to its standard expectation of $A_{\rm L}=1$ when this parameter is freely varying. However in this case this parameter is fixed and therefore a reduced growth of structure can mimic the very same effect. Consequently, the value of the growth index $\gamma$ approaches the standard one when including CMB lensing in the data analyses, i.e., shifts to larger values, getting closer to $\gamma=0.55$. As can be noticed from Tab.~\ref{tab:gamma}, the values of  the Hubble parameter are not significantly shifted and therefore this simplest model does not ameliorate the so-called 
 $H_0$ tension. The values of the $S_8$ parameter are instead lower than within the minimal $\Lambda$CDM cosmology and closer to those obtained from weak lensing probes, alleviating therefore the so-called $\sigma_8$ tension. Indeed, the mean values of $S_8$ are very similar to those reported when combining CMB data with observations from the Dark Energy Survey (DES). The difference within this simplest model in the values of $\gamma$ obtained here and those reported in Ref.~\cite{Nguyen:2023fip} can be due to a number of differences among the two analyses. Namely, here we modify the evolution of the two gravitational potentials rather than simply change the matter power spectrum evolution. Also, we do not make use of datasets which may require the use of the halofit numerical code which has not been properly modified to account for departures from the canonical growth of structure. As previously argued, this is the main reason to not to exploit weak lensing observations in almost all the results shown here.  Figure~\ref{fig:comparison} shows the fact that, if we include the very same observations than those considered in Ref.~\cite{Nguyen:2023fip}, we get the opposite to what we report here, i.e. a reduced growth factor characterized by $\mu_0-1<0$, or, equivalently, by $\gamma>0.55$. The additional data sets considered in Fig.~\ref{fig:comparison} are the Dark Energy Survey 3x2 point correlation functions, referred to as DESY1~\cite{DES:2017myr,DES:2017gwu,DES:2017hdw,DES:2017qwj,DES:2018ufa} and measurements of $f\sigma_8$, being $\sigma_8$ the clustering parameter,  extracted from both peculiar velocity and redshift-space distortion (RSD) data~\cite{Said:2020epb,Beutler:2012px,Huterer:2016uyq,Boruah:2019icj,Turner:2022mla,Blake:2011rj,Blake:2013nif,Howlett:2014opa,Okumura:2015lvp,Pezzotta:2016gbo,eBOSS:2020yzd}.  Specifically, we use the so-called ``Gold-2017'' growth-rate dataset~\cite{Nesseris:2017vor}. We refer this dataset as "RSD".

 Table~\ref{tab:mueta} shows the mean values and the errors obtained via the parameterization given by Eq.~(\ref{eq:mueta}). Notice that for all the data  combinations there are not significant departures from the standard cosmological expectations $\mu_0=\eta_0=1$, but, if any, \textit{they always imply a larger growth, i.e. $\gamma <0.55$} and not a reduced one, as claimed by Ref.~\cite{Nguyen:2023fip}  due to a partial parameterization of the growth effects, only included in the matter power spectrum. The situation with the $H_0$ and $\sigma_8$ tensions is very similar to that observed within the $\gamma$-only parameterization. Nevertheless, the value of the parameter $S_8$ is further reduced in this case, improving therefore the consistency on the values of this parameter obtained by CMB observations and by weak lensing probes. 
 Table~\ref{tab:gammaeta} shows the mean values and the errors obtained via the parameterization given by Eq.~(\ref{eq:gamma}) and the second of Eqs.~(\ref{eq:mueta}).  Notice that from CMB data alone there is a $2\sigma$ preference for $\gamma<0.55$. In this case, when we add lensing observations, the significance is improved to the $3\sigma$ level, but the mean value of $\eta_0$ gets closer to 1. Therefore, as long as CMB lensing observations are considered, there is a $3\sigma$ preference for a growth enhancement, while if these observations are not considered,  such a preference is reduced to the $2\sigma$ level. 

\subsubsection{CMB lensing amplitude $A_{\rm L}$}
The values of $\gamma$ and $\eta$ reported above clearly show that there is a preference for an enhanced growth of structure from CMB observations, either alone or combined with BAO and SNIa luminosity distances. The larger growth that we obtain at late times would increase CMB lensing and therefore imply another reflection of the preference 
for a lensing amplitude $A_{\rm L} > 1$. Consequently, we perform  also here analysis in which this parameter is a freely varying one. 
Table \ref{t1} shows that the extension of the $\Lambda$CDM model with $A_{\rm L}$ is moderately disfavoured with respect to the minimal scenario. More importantly here, the ($\gamma$, $\eta$) plus $A_{\rm L}$ case is weakly disfavoured since, as we have already discussed, the $\gamma$ parameter and the lensing amplitude have the very same effect in the lensing observables due to large scale structure effects. Allowing $A_\ell$ to be an oscillatory function of the multipole $A_{\rm L} + A_m \sin \ell$ is a weakly or moderately disfavoured scenario, either within the minimal $\Lambda$CDM cosmology or within the ($\mu$, $\eta$) parameterization. 

\begin{figure*}
\begin{tabular}{c c}
 \includegraphics[width = 0.5\textwidth]
{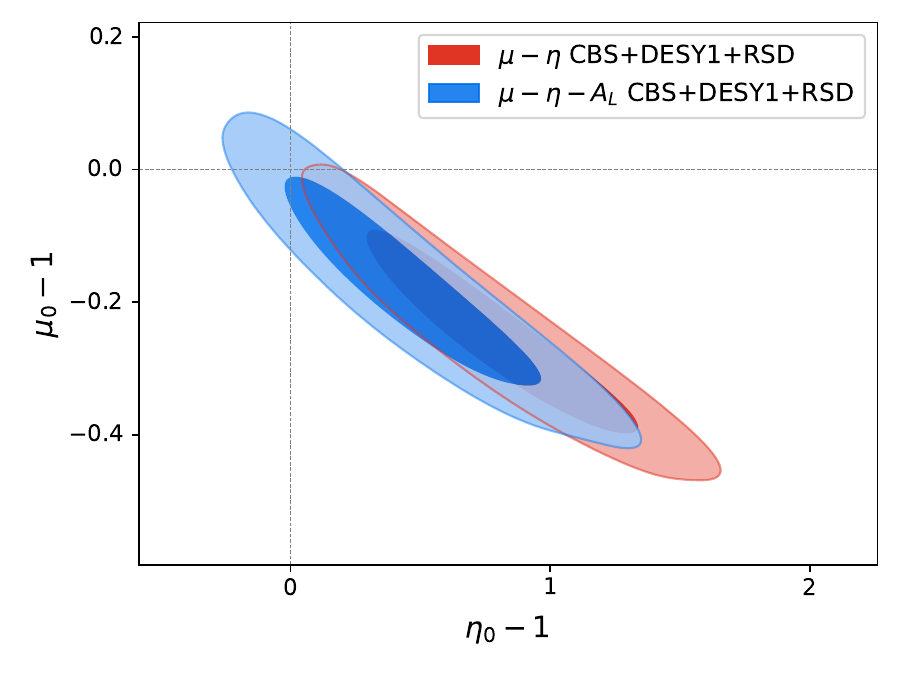} &
\includegraphics[width = 0.5\textwidth]
{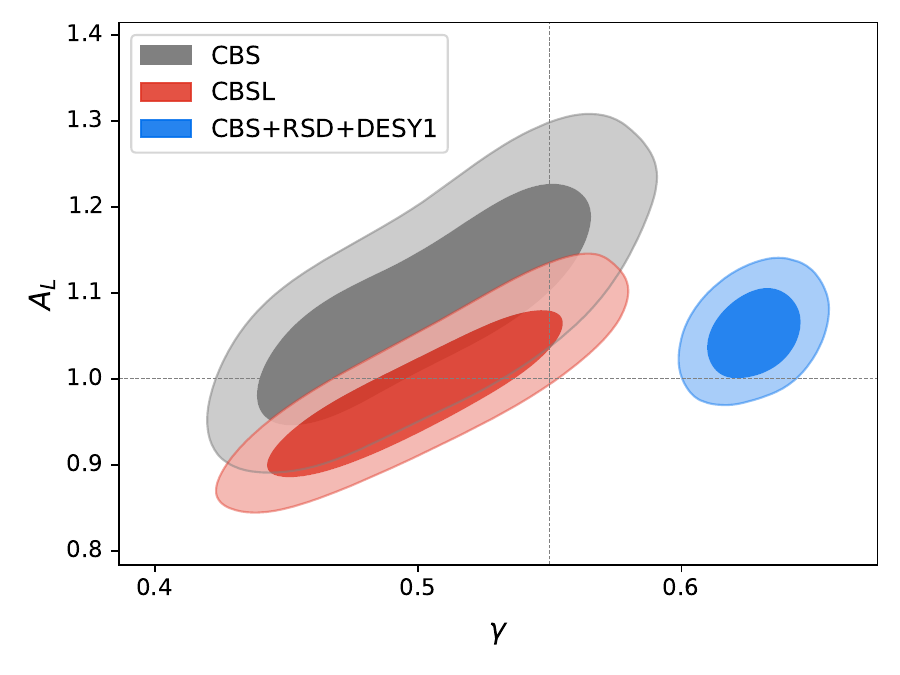} 
\end{tabular}
\caption{\textit{Left panel:} The red regions depict the $68\%$ and $95\%$ CL contours in the ($\eta_0-1$, $\mu_0-1$) plane within the growth model parameterized via $\mu$ and $\eta$, see  Eqs.~(\ref{eq:mueta})  obtained by using CMB temperature and polarization measurements, BAO, SNIa, weak lensing and RSD measurements. The  blue regions depict the equivalent  but when the lensing amplitude $A_L$ is also considered as a freely varying parameter. \textit{Right panel:}. Allowed regions in the simplest growth parameterization used here, i.e. via just one single parameter $\gamma$ in the ($\gamma$, $A_{\rm L}$) plane allowed from different data sets, for the sake of comparison with the results of Ref.~\cite{Nguyen:2023fip}. Thee horizontal and vertical lines depict the expectations within the canonical $\Lambda$CDM cosmology, i.e. $A_{\rm L}=1$ and $\gamma=0.55$, respectively.}
\label{fig:comparison}
\end{figure*}

The right panels of Figs.~\ref{fig:mueta} and \ref{fig:gammaeta} depict the 68\% and 95\%~CL allowed contours from  Planck CMB data, with and without CMB lensing, in the ($\mu_0-1$, $\eta_0$-1) and ($\gamma$, $\eta_0$-1) planes, associated to the  ($\mu$, $\eta$) and ($\gamma$, $\eta$) parameterizations respectively, with $A_{\rm L}$ a freely varying parameter in the data analyses. Notice that the mean values of $\gamma$ are now shifted towards its expectation within the $\Lambda$CDM cosmology, $\gamma=0.55$. 
Figure~\ref{fig:comparison} (left panel) shows that the  reduced growth of structure favoured by weak lensing and RSD observations is less evident when adding $A_{\rm L}$ as a free parameter. The presence of a varying lensing amplitude restores the canonical values for 
($\mu_0-1$, $\eta_0-1$), i.e. shifts their values towards $(0,0)$.  The right panel of Fig.~\ref{fig:comparison} clearly illustrates the large degeneracy between $A_{\rm L}$ and $\gamma$ within the simplest growth model explored here, i.e. the one with one single parameter $\gamma$. Notice that the effect of a larger lensing amplitude $A_{\rm L}$ can be compensated with a smaller growth of structure, i.e. a value of $\gamma$ closer to the standard expectation of $\gamma=0.55$ in the case of the CBS and CBSL data sets. When considering also RSD and DESY1 measurements the mean value of $\gamma>0.55$ but the very same degeneracy holds, i.e., larger lensing amplitudes are correlated with a smaller growth of structure ($\gamma>0.55$).

\begin{figure*}
\begin{tabular}{c c}
 \includegraphics[width = 0.5\textwidth]
{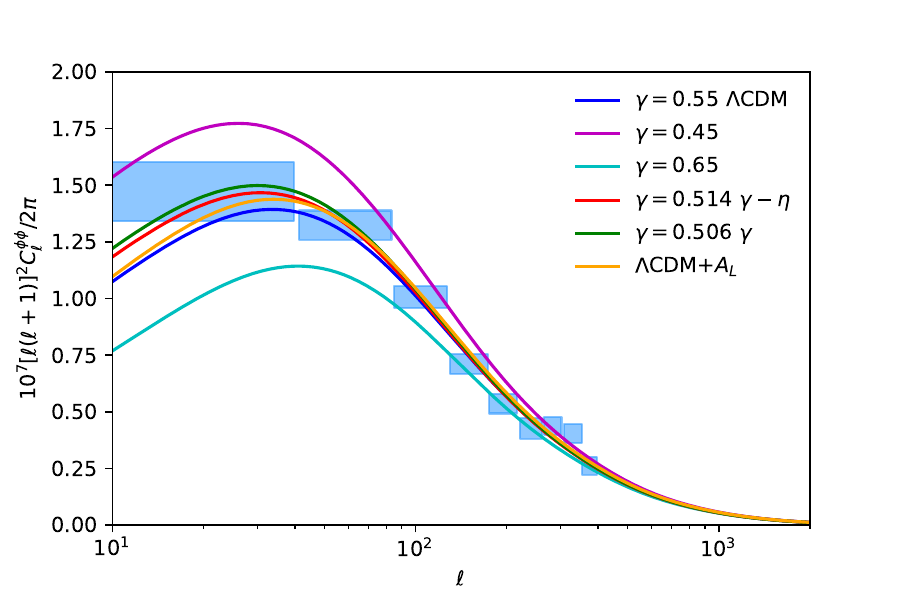} & 
 \includegraphics[width = 0.5\textwidth]
{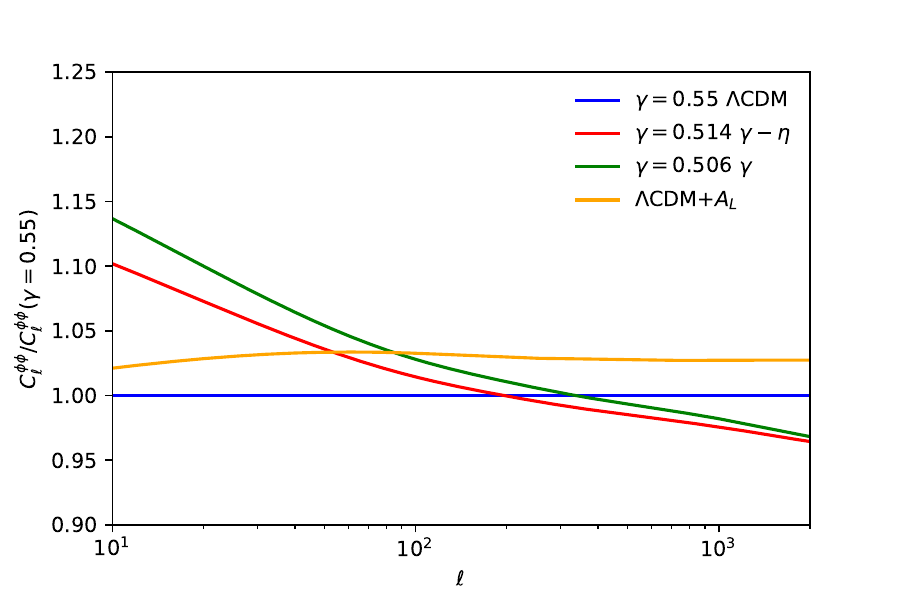} 
\end{tabular}
\caption{\textit{Left panel:} CMB lensed power spectrum, together with Planck 2018 measurements for several possible scenarios, see text for details.  \textit{Right panel:} Deviation of the lensing potential with respect to the $\Lambda$CDM case for different growth parameterizations.}
\label{fig:lensing}
\end{figure*}

The fact that both $\gamma$ and $A_{\rm L}$ have a very similar impact in the CMB lensing power spectra at late times is also explicitly shown in Fig.~\ref{fig:lensing}. The left panel shows the CMB lensed power spectrum, together with Planck 2018 observations~\cite{Planck:2018lbu} for several possible scenarios. In the case of $\Lambda$CDM plus $A_{\rm L}$, we have fixed the former parameter to $1.07$, which is the best fit from Planck temperature, polarization and lensing data. Notice that, if $\gamma<0.55$, the lensing power spectrum is enhanced, due to a larger growth of structure. The  very same enhancement is obtained by a value of $A_{\rm L} >1$. The right panel of  Fig.~\ref{fig:lensing} depicts the deviation of the lensing potential with respect to the $\Lambda$CDM case for the different growth parameterizations exploited here. Notice that for multipoles $\ell<200$ the lensing power spectrum is larger than that expected within the canonical scenario if either $\gamma<0.55$ or $A_{\rm L}>1$.

Table~\ref{tab:lcdmal} shows the preferred values for the lensing amplitude $A_{\rm L}$ within the minimal $\Lambda$CDM universe from the different data  combinations considered here. While CMB data (either alone or in combination with BAO and SNIa measurements) prefers $A_{\rm L}>1$ with $\sim 3 \sigma$ significance, the addition of CMB lensing observations softens this preference to the $2\sigma$ level.

Table~\ref{tab:gammaal}  depicts the equivalent to Tab.~\ref{tab:gamma} but leaving $A_{\rm L}$ as a free parameter.  
Notice that we obtain values of $\gamma<0.55$ always, but the statistical significance for such a preference is decreased due to the fact that $\gamma$ is degenerate with the lensing amplitude, i.e. a lower value of $\gamma$ can be compensated with a lower value of $A_{\rm L}$ and also because lensing observations prefer a lower value of the lensing amplitude.  The preference for $\gamma<0.55$ decreases to the $1-2\sigma$ significance level, depending on the data sets. Furthermore, when CMB lensing observations are considered, the mean value of $A_{\rm L}<1$.

Table~\ref{tab:muetaal} shows the equivalent to Tab.~\ref{tab:mueta} but when $A_{\rm L}$ is a free parameter, i.e. the mean values and the errors obtained via the parameterization given by Eq.~(\ref{eq:mueta}) plus a freely varying lensing amplitude parameter. As in the $A_{\rm L}=1$ case, for all the data  combinations explored here there are not significant departures from the standard cosmological expectations $\mu_0=\eta_0=1$, but, if any, \textit{they always imply a larger growth, i.e. $\gamma <0.55$} and not a reduced one. As in the previous case we no longer observe a significant preference for $A_{\rm L}>1$, being this parameter smaller than 1 when CMB lensing observations are included in the data analyses. This is related to the degeneracy between $A_{\rm L}$ and the $(\mu, \eta)$ parameterization and also to the fact that CMB lensing measurements prefer a lensing amplitude closer to the standard value $A_{\rm L}=1$. 

Table~\ref{tab:gammaetaal} shows the equivalent to Tab.~\ref{tab:gammaeta} but leaving $A_{\rm L}$ as a free parameter, that is, presents the results in the parameterization given by Eq.~(\ref{eq:gamma}) and the second of Eqs.~(\ref{eq:mueta}). In this case, there is always a  $2\sigma$ preference for $\gamma<0.55$. As aforementioned, the preference for $A_{\rm L}>1$
is very mild, being the mean value of this parameter
$A_{\rm L}<1$ when CMB lensing observations are considered.

The last possibility we explore here is to consider the lensing amplitude to be an oscillatory function of the multipole $\ell$, $A_\ell= A_{\rm L} + A_m\sin  \ell$.  
 Table~\ref{tab:lcdmosc} shows the results for $A_{\rm L}$ and $A_m$ within the minimal $\Lambda$CDM cosmology. Notice that the data show no particular preference for $A_m\neq 1$. A value of $A_{\rm L} > 1$ is still favoured with $2\sigma$ significance. Table~\ref{tab:muetaosc} shows the equivalent but for the ($\mu$, $\eta$) parameterization. While no strong preference for $\mu\neq 0$ and $\eta \neq 0$ is found from any of the data combinations, the mean values of $\mu_0$ and $\eta_0$ are always  larger than 1, shifting the standard growth of structure towards an \emph{enhanced value, i.e. $\gamma< 0.55$}.

A final comment should be devoted to the different data sets exploited here. Notice that the CBS combination is the most conservative and ideal one, and, since, as it contains less data sets, it  can avoid potential uncertainties present in the data. Also the statistical significance for $\gamma>1$ is in general larger than when considering lensing. 
The combination ``CBSL'', being more complete, decreases the significance of the signal and also the preference for $A_{\rm L}>1$. Nevertheless, it introduces more uncertainties in the analyses, as it relies on the halofit numerical code, which has not been properly updated in the context of the different growth parameterizations considered here.

\begin{figure}
	\centering
	\includegraphics[scale=0.55]{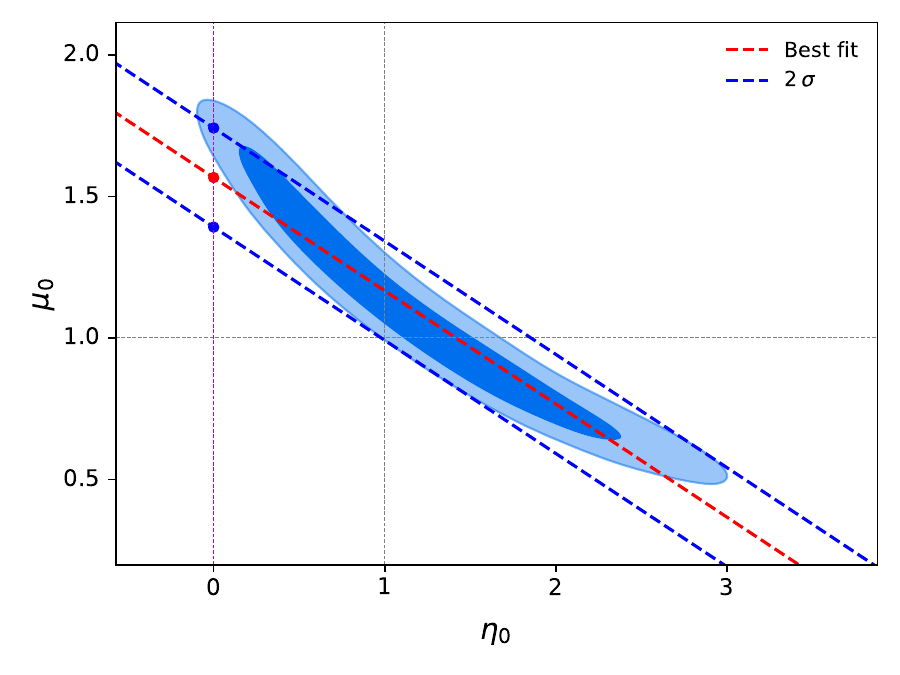}
	\caption{$S_0$ acts as a signal parameter to measure the deviation from the standard cosmic growth in general relativity. The red and blue points denote the best fit value and $2\sigma$ limits of $S_0$, respectively. Similarly, the red and blue lines are the best fit line and $2\sigma$ boundaries when using the fitting formula $S_0=\mu_0+0.4\eta_0$, respectively. The magenta dashed line corresponds to $\eta_0=0$ and the cross point between grey dashed lines represents general relativity. The blue contours depict the $68\%$ and $95\%$ confidence regions in the ($\eta_0$, $\mu_0$) plane within the growth model parameterized via $\mu$ and $\eta$, see Eqs.~(\ref{eq:mueta}), obtained by using CMB temperature, polarization and lensing observations.}\label{S0}
\end{figure}

\begin{figure*}
	\centering
	\includegraphics[scale=0.7]{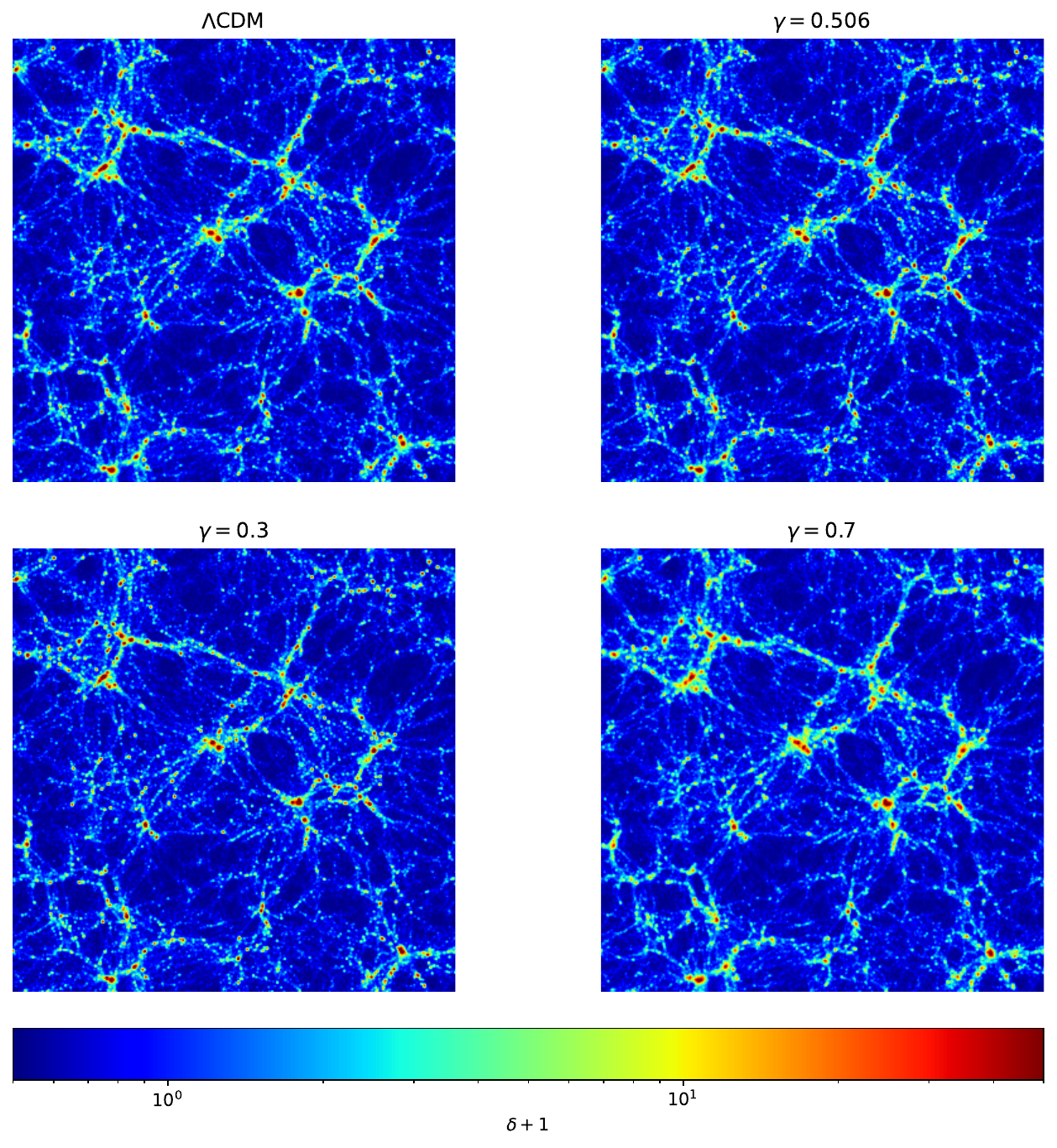}
	\caption{Density fields of dark matter in the growth index $\gamma$ single parameter model for the values $\gamma=0.506$, $0.3$ and $0.7$ at redshift $z=0$. For the case of $\gamma=0.506$, we use the best fit of the growth index model to implement the simulation. We also depict the $\Lambda$CDM case as a reference and choose a slice of 20 $h\,\mathrm{Mpc^{-1}}$ density field and stack it along $Z$ axis for each model. $\delta$ denotes the overdensity of dark matter.}\label{f1}
\end{figure*}

\begin{figure*}
	\centering
	\includegraphics[scale=0.55]{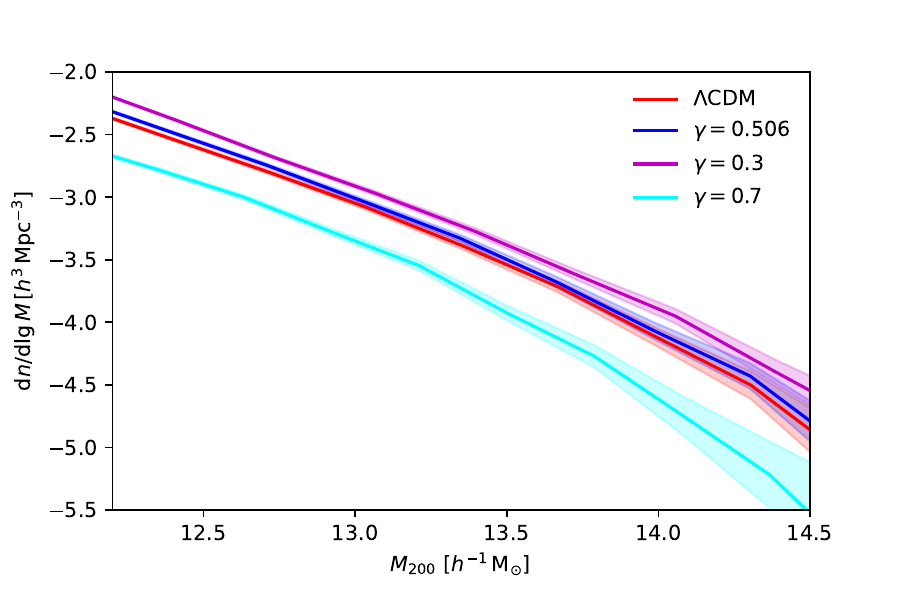}
	\includegraphics[scale=0.55]{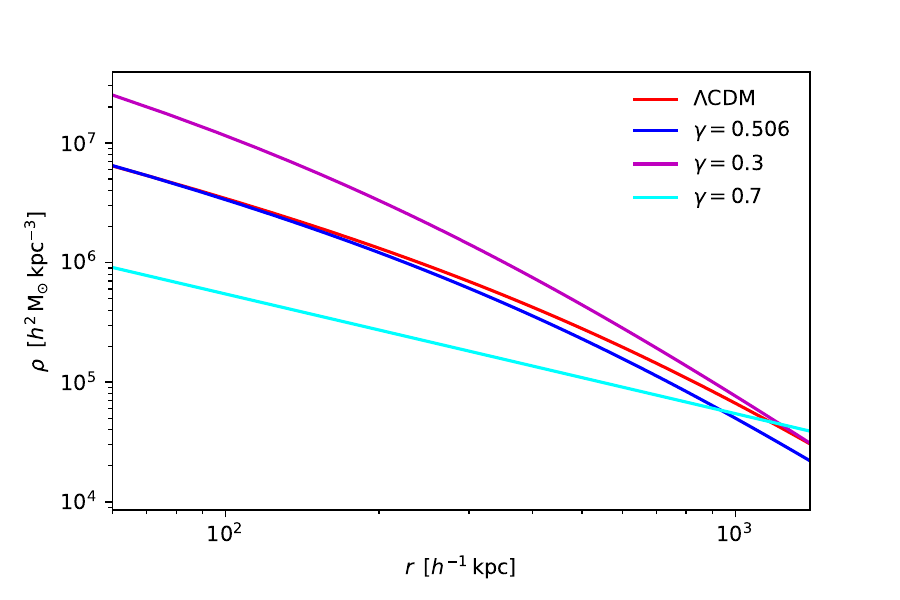}
	\caption{{\it Left panel:} The dark matter halo mass functions in the growth index model are, respectively, shown for different values $\gamma=0.506$, $0.3$ and $0.7$ at redshift $z=0$. We also depict the halo mass function of the $\Lambda$CDM case as a reference. The shaded regions denote the Poisson errors for each model. {\it Right panel:}  The dark matter halo density profiles for different models are shown.}\label{f2}
\end{figure*}

\subsubsection{Signal parameter $S_0$ measuring cosmic growth}
In the above analysis, although we have shown the constraints on different growth parameterizations from different data combinations, there is still an important and interesting issue to be addressed, i.e., what does $\gamma$ or $(\mu_0, \eta_0)$ test at all? For example, one can easily find that CMB plus lensing gives $\gamma=0.506\pm0.022$ in the growth index model, which is a $2\sigma$ signal of enhanced structure growth (see Tab.\ref{tab:gamma}), and $2\sigma$ deviation from general relativity in the $(\mu_0, \eta_0)$ plane (see \ref{fig:mueta} and Fig.\ref{S0}). Indeed, the information of this enhanced growth can be captured and described accurately by a linear combination of $\mu_0$ and $\eta_0$, namely by the approximate fitting formula $S_0=\mu_0+0.4\eta_0$, where $S_0$ denotes the intercept in the $\mu_0$ axis when $\eta_0=0$. Using the data combination of CMB temperature, polarization and lensing, we obtain the constraint $S_0=1.567\pm0.088$ at $2\sigma$ level. Note that $S_0$ is a derived parameter here. In Fig.~\ref{S0}, we have presented how $S_0$ serves as a signal parameter to measure the deviation from the standard growth in general relativity. It is easy to see that the $S_0$ parameter captures the $2\sigma$ excess very well. This means that $S_0$ encodes what $\gamma$ or $(\mu_0, \eta_0)$ tests. This parameter extracts efficiently growth information which is hiding in both effective gravitational strength and anisotropic stress of cosmic species. $S_0$ can be generalized to test the deviation from general relativity in large scale structure observations.

\subsubsection{N-body simulations}

Using the halofit model, one can easily express the nonlinear matter power spectrum as the one-halo plus the two-halo terms
\begin{equation}
P_{\mathrm{nonlinear}}(\gamma,k,a) = P_{\mathrm{1h}}(\gamma,k,a)+P_{\mathrm{2h}}(\gamma,k,a),
\end{equation}
where one-halo term is
\begin{equation}
P_{\mathrm{1h}}(\gamma,k,a) = \int_0^\infty W^2(\gamma,M,k,a)n(\gamma,M,a)\mathrm{d}M~,
\end{equation}
and two-halo term reads
\begin{widetext}
\begin{equation}
P_{\mathrm{2h}}(\gamma,k,a) = P(\gamma,k,a)\left[\int_0^\infty W(\gamma,M,k,a)n(\gamma,M,a)b(\gamma,M,a)\mathrm{d}M\right]^2~,
\end{equation}
\end{widetext}
where $W(\gamma,M,k,a)$, $n(\gamma,M,a)$ and $b(\gamma,M,a)$ are halo density profiles, halo mass function and halo bias, respectively. It is easy to see that these three halo related quantities all depend on $\gamma$, halo mass $M$ and $a$. Note that the halo density profile also depends on the scale $k$.

\begin{table}[!t]
	\renewcommand\arraystretch{1.5}
	\caption{The expected galaxy number densities and biases for SKA2 at the redshift range of interest are shown. Note that the galaxy number densities are in units of Mpc$^{-3}$.}
	\setlength{\tabcolsep}{6mm}{
	\begin{tabular} { c | c | c }
		\hline
		\hline
       $z$  & $n_(z)\times10^{-6}$  & $b(z)$  \\
		\hline
		  0.23    & 44300    & 0.713            \\
		  0.33    & 27300   &   0.772          \\
		0.43    & 16500   &   0.837          \\
		0.53    & 9890   &    0.907         \\
		0.63    & 5880   &    0.983         \\
		0.73    & 3480   &    1.066         \\
		  0.83    & 2050   &    1.156        \\
		  0.93    & 1210   &    1.254         \\
		  1.03    & 706   &     1.360        \\
		1.13    & 411   &     1.475        \\
		1.23    & 239   &     1.600        \\
		  1.33    & 139   &     1.735        \\
		  1.43    & 79.9   &    1.882         \\
		1.53    & 46.0   &    2.041         \\
		1.63    & 26.4   &    2.214         \\
		1.73    & 15.1   &    2.402         \\
		  1.81    & 9.66   &    2.566         \\

		\hline
		\hline
	\end{tabular}
	\label{tab:bias}}
\end{table}

The authors of Ref.~\cite{Nguyen:2023fip} claim that the linear power spectrum enters only through the variance of the matter density field in the halo mass function $n(M,\,a)$. In light of this setting, they derive their cosmological constraints by using weak lensing and RSD observations. However, the realistic consequence is that $\gamma$ also affects both $W(\gamma,M,k,a)$ and $b(\gamma,M,a)$ in a completely nonlinear way. These effects need to be captured by numerical simulations. Therefore, based on this concern, in order to explore this issue more accurately, we implement dark matter simulations  (exclusively for the simple growth index one single parameter model). In Fig.\ref{f1} we show the density fields of dark matter for the growth index model with different values. Namely, we illustrate the cases $\gamma=0.506$ (see Tab.\ref{tab:gamma}), $0.3$ and $0.7$, respectively. It is easy to notice that the density fields of both $\Lambda$CDM and the best fit growth index model ($\gamma=0.506$) are very close. However, for the case of $\gamma=0.3$, one can observe a significant enhancement of structure growth relative to the $\Lambda$CDM case at cluster scales. Since gravitational strength is stronger than $\Lambda$CDM everywhere, a value of $\gamma=0.3$ accelerates the process of structure formation. It is noteworthy that we do not clearly observe more small structures than in $\Lambda$CDM. For the case of $\gamma=0.7$, the result is almost opposite to the case of $\gamma=0.3$. We find weaker clustering of dark matter at both large and small scales and lower overdensities relative to the $\Lambda$CDM case due to weak gravitational strength: similar structures would need more time to form when compared to the $\Lambda$CDM case. 

Furthermore, we provide a quantitative analysis of different simulations. Specifically, in Fig.\ref{f2}, we present the dark matter halo mass functions and halo density profiles for different values of $\gamma$ in the growth index model.  In light of the CMB plus lensing constraint, one can easily find that the best fit $\gamma$ model predicts a little larger number of halos than $\Lambda$CDM does in the whole halo mass range. We are also interested in investigating the effects of large and small $\gamma$ values on the cosmic web. In the case of $\gamma=0.3$, an obvious enhancement of structure formation is observed relative to $\Lambda$CDM, since there are more halos generated for all the halo masses. On the contrary, fewer halos are produced in the case of $\gamma=0.7$, which indicates that structure growth is clearly suppressed for a small $\gamma$ value. In general, these statistics properties of the halo number can be seen in the density fields of dark matter (see also Fig.\ref{f1}). 

A representative statistical quantity describing the dark matter halo, the density profiles of dark matter halos, are also computed for the different models. Note that, for simplicity, we only study the density profile of the largest halo for each model here. We find that the best fit $\gamma$ model shows the same density as $\Lambda$CDM when the radius $r<100\,h^{-1}\mathrm{kpc}$. Nonetheless, it gives a lower density when $r$ becomes large (say $1000\,h^{-1}\mathrm{kpc}$), and consequently predicts a lower average density of the whole halo. In the case of $\gamma=0.3$, the halo density is clearly stronger than that in $\Lambda$CDM in the whole range of radius. Interestingly, the slope of its density profile becomes larger and reaches the same value when $r>1000\,h^{-1}\mathrm{kpc}$ relative to $\Lambda$CDM. This implies that, in a universe with an enhanced structure formation, both the average density of a halo and the halo compactness are larger than that in a $\Lambda$CDM one. In the case of $\gamma=0.7$, the halo density is suppressed at a small radius but becomes larger than other cases starting from $r\simeq1000\,h^{-1}\mathrm{kpc}$. This behavior is due to the fact that a weak structure growth strength decelerates the structure formation of the universe and consequently decreases the matter accretion process of the halo. All in  all,  the growth index $\gamma$ has an important impact on the evolution of cosmic web.

\begin{figure}
	\centering
	\includegraphics[scale=0.55]{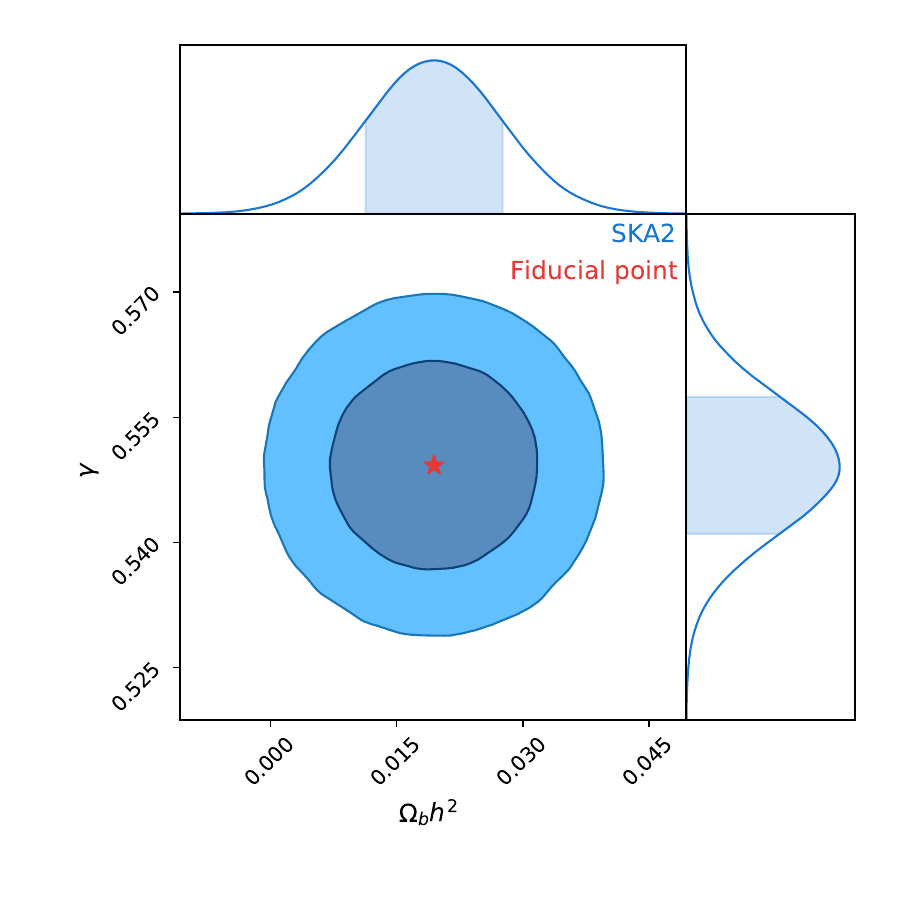}
	\caption{The cosmological forecasts from the future HI galaxy redshift survey SKA2 is shown in the ($\Omega_bh^2$,\,$\gamma$) plane. }\label{f3}
\end{figure}

\begin{table}[!t]
	\renewcommand\arraystretch{1.5}
	\caption{Predicted uncertainties from SKA2 and measured errors from Planck-2018 CMB data.}
	\setlength{\tabcolsep}{6mm}
        {\begin{tabular}{@{}ccc@{}} \toprule
			Parameters            &SKA2     &Planck      \\ \colrule
		    $\sigma(\gamma)$      &0.0083   &0.022       \\  
                $\sigma(H_0)$         &0.19     &0.54       \\
                $\sigma(\sigma_8)$    &0.00079  &0.006        \\
			\botrule
		\end{tabular}
		\label{tab:forecast}}
\end{table}

\subsubsection{Forecasts for future 21 cm surveys}
In the near future, there will be two type of (crucial) 21 cm large scale structure probes: 21 cm intensity mapping and HI galaxy redshift surveys \cite{Ansari:2022nmy}. The fluctuations of the brightness temperature induced by redshifted 21 cm lines trace the HI distribution and consequently detect the large scale structure of the universe. HI galaxy redshift surveys allow us to measure the cosmic expansion history using BAO as well as the cosmic structure growth using RSD. One can constrain the matter power spectrum or correlation functions by identifying individual galaxies and confirming their redshifts, and consequently determine the cosmological parameters. During the past two decades, optical galaxy redshift surveys such as SDSS \cite{eBOSS:2020yzd,Ross:2020lqz,sdss} have achieved great success in exploring the universe. We believe that future high-resolution HI galaxy surveys such as the Square Kilometre Array (SKA) \cite{ska,Santos:2015gra} can give stronger constraining power on both the cosmic geometry and the growth. In this study, we focus on the forecasted constraints on the different cosmological parameters by employing the HI galaxy redshift data from SKA Phase 2 (hereafter SKA2).  

The galaxy power spectrum for a galaxy redshift survey is expressed as
\begin{equation}
P_g(\gamma, k,z)=\left[b(z)+f\xi^2\right]^2e^{-\frac{\left[k\sigma_{\mathrm{NL}}(z,\xi)\right]^2}{2}}P(\gamma, k, z),
\label{eq:GPS}
\end{equation} 
where the first term depicts the so-called Kaiser effect \cite{Kaiser:1987qv}, the second one describes the ``Finger of God'' effect due to uncorrelated velocities at small scales, which washes out the radial structure below the nonlinear velocity dispersion scale $\sigma_{\mathrm{NL}}$, $\xi\equiv\hat{k}\cdot\hat{z}$, $b(z)$ is the galaxy bias as a function of redshift $z$ and 
\begin{equation}
\sigma_{\mathrm{NL}}(z,\xi)=\sigma_{\mathrm{NL}}D(z)\sqrt{1+f\xi^2(2+f)}.
\label{eq:dispersion}
\end{equation}  
The quality of a galaxy redshift survey is subject to complex systematic effects. In particular, there are two main effects, namely the source evolution and bright stars as contaminants. For the former case, the luminosity function of the tracer population usually changes over redshift, which affects the detected number density of galaxies. This effect,  generally characterized by the galaxy bias $b(z)$, limits the effective redshift range of a galaxy survey and make its selection function more complicated. It affects the measurement of BAO by varying the effective galaxy number density $m(z)$ to change the shot noise. It is worth noting that stars are a dominated contaminant in large optical galaxy sky surveys. Bright stars can mask galaxies behind them, when one distinguishes stars from galaxies by their color. This effect leads to a very complicated angular selection function on the sky. It is interesting that this effect is not very dangerous in the radio wavelength where SKA works, even though there are also other contaminants such as diffuse galactic synchrotron emission and some non-galaxy point sources affecting the final galaxy catalogue and source-finding process, see Ref.~\cite{Bull:2014rha} for further details. 

Instead of producing mock data by implementing a full simulation, we adopt a fast and low computational cost method to forecast the uncertainties on cosmological parameters, i.e., the Fisher Matrix formalism, which transforms the expected properties of signal and noise in theoretical quantities for a given survey to derive a Gaussian approximation into the underlying likelihood for a set of parameters to be measured. This method plays an important role in characterizing the ability of a given experiment to constrain the parameters of interest.

For a set of parameters $\mathbf{p}$ given the data $\mathbf{d}$, we assume that the  likelihood function $\mathcal{L}$ is a Gaussian distribution and can be written as 
\begin{equation}
\mathcal{L}(\mathbf{p}|\mathbf{d})\propto\frac{\mathrm{exp}\left(-\frac{1}{2}\mathbf{d}^\dagger\left[\mathbf{M}(\mathbf{p})\right]^{-1}\mathbf{d}\right)}{\sqrt{|\mathbf{M}(\mathbf{p})|}}~, 
\label{eq:like}
\end{equation}
where $\mathbf{M}$ is the covariance matrix of the mock data. The information of cosmological parameters seeds in the parameter vector $\mathbf{p}$. Using the fiducial values of parameters, the Fisher Matrix can be constructed by the curvature of the likelihood function as
\begin{equation}
F_{\alpha\beta}\equiv-\left<\frac{\partial^2 \mathrm{log}\mathcal{L}}{\partial p_\alpha\partial p_\beta}\right>_{\mathbf{p=p_0}}, \label{eq:fisher}
\end{equation}
where $\mathbf{p_0}$ denotes the fiducial parameter vector. 

For a galaxy redshift survey, the Fisher matrices read as 
\begin{equation}
F_{\alpha\beta}=\frac{1}{2}\int\frac{\mathrm{d}^3k}{(2\pi)^3}V_{\mathrm{eff}}(k)\left(\frac{\partial \mathrm{ln\,S}_T}{\partial \, p_\alpha}\frac{\partial \mathrm{ln\,S}_T}{\partial \, p_\beta}\right), \label{eq:fishergalaxy}
\end{equation}
where $S_T$ is the total covariance of the measured signal, which consists of the underlying signal $S_S$ and the noise $S_N$, while $V_{\mathrm{eff}}(k)=V(S_S/S_T)^2$ denotes the effective volume of the experiment covering a physical volume $V$.  Two parts contribute to the total signal: the galaxy power spectrum (see Eq.(\ref{eq:GPS})) and the shot noise, namely $S_T=P(k,z)+1/m(z)$.

SKA2 \cite{Santos:2015gra} will be very sensitive and it is expected to achieve an root-mean-square (rms) flux sensitivity $S_{\mathrm{rms}}\approx5 \, \mathrm{\mu Jy}$ covering a sky area of 30000 deg$^2$ for 10000 hours. SKA2 aims to produce a catalogue of one billion HI galaxies in the redshift range $z\in(0.18, \, 1.84)$, which is far beyond any planned optical or near infrared experiments when $z\in(0, \, 1.4)$. The expected galaxy number densities and galaxy biases for SKA2 are shown in Tab.~\ref{tab:bias}. 

After numerical computations, the forecasted results from SKA2 are shown in Fig.~\ref{f3} and Tab.~\ref{tab:forecast}. We obtain the $1\sigma$ uncertainty $\sigma(\gamma)=0.0083$ on the growth index $\gamma$, i.e. a 1.5\% determination of the cosmic structure growth, assuming a  fiducial value $\gamma=0.55$ and therefore reducing the current errors by a factor of three. Similarly, we obtain the constraint $\sigma(H_0)=0.19$ km~s$^{-1}$~Mpc$^{-1}$ relative to $H_0=67.36$ km~s$^{-1}$~Mpc$^{-1}$, which provides a 0.3\% determination on $H_0$ and increases the accuracy of Planck's measurement by a factor of three. Interestingly, we find the $1\sigma$ error of matter clustering amplitude $\sigma(\sigma_8)=0.00079$, which implies a 0.098\% prediction of $\sigma_8$ and improves Planck's accuracy by a factor of eight.

\section{Discussions and conclusions}
\label{sec:conclusions}
A number of cosmological tensions suggest the necessity of testing the canonical cosmological $\Lambda$CDM paradigm at a variety of scales. Some of these tensions may require to modify the standard model of structure formation in our universe. It is therefore timely to evaluate whether current cosmological observations point to a departure from the standard growth picture. In the simplest parameterization, the growth of structure is simply modeled via a single parameter $\gamma$, with $f\equiv d\delta/d\ln a \equiv \Omega_m(a)^\gamma$ and $\gamma \simeq 0.55$ in the standard $\Lambda$CDM case. A recent analysis has suggested a lower structure growth which in turns implies $\gamma>0.55$. Here we further explore this issue extending the analysis to other possible growth parameterizations. In all the cases, for the set of cosmological observations considered here, we obtain \emph{a higher growth of structure}, characterized by $\gamma < 0.55$. Such a preference reaches the $3\sigma$ significance using Cosmic Microwave Background observations, Supernova Ia and Baryon Acoustic Oscillation measurements, while the addition of Cosmic Microwave Background lensing data relaxes such a preference to the $2\sigma$ level, since a larger lensing effect can always be compensated with a smaller structure growth, or, equivalently, with $\gamma>0.55$. Nevertheless, when considering the very same data sets that those previously considered in the literature regardless the validity of the analyses based on assumptions on the halo model~\cite{Nguyen:2023fip}, we are also able to reproduce previous claims. However, the most conservative data combination considered throughout our analyses and the self-consistent growth parameterization employed here 
make the robust prediction of \emph{an enhanced structure growth}, rather than a suppressed one. We have also included the lensing amplitude $A_{\rm L}$ as a free parameter in our data analyses, showing that the preference for $A_{\rm L}>1$ observed within the canonical $\Lambda$CDM paradigm still remains, except for some particular parameterizations when lensing observations are included.
 There is a clear very large degeneracy between the lensing amplitude and the growth rate index $\gamma$, as a larger value of $A_{\rm L}$ implying more lensing can always be compensated by $\gamma>0.55$, reducing structure formation and also CMB lensing (to compensate for the effect of a larger $A_{\rm L}$). 
We also not find any significant preference for a multipole dependence of $A_{\rm L}$. To further reassess the effects of a non-standard growth index, we have computed by means of N-body simulations the dark matter density field,  the
dark matter halo mass functions and the halo density profiles
for different values of $\gamma$ in the growth index model. For $\gamma <0.55$, a large number of halos and a stronger dark matter halo density profile are found, with respect to the canonical $\Lambda$CDM case. Devoted forecasted analyses by means of the Fisher matrix methodology have also been performed in order to study the future sensitivity of the SKA HI galaxy redshift data to the growth index $\gamma$, showing that the current errors could be reduced by a factor of three, testing further the validity of the standard model of structure formation.

\begin{table*}[!t]
	\renewcommand\arraystretch{1.6}
	\caption{Mean values and $68\%$~CL marginalized errors of the most relevant cosmological parameters within the growth model parameterized via $\gamma$, obtained by using the CMB plus lensing, CMB alone, CMB plus BAO plus Supernovae Ia plus lensing and CMB plus BAO plus Supernovae Ia datasets, respectively.}
	\setlength{\tabcolsep}{6mm}{
	\begin{tabular} { l |c| c |c| c }
		\hline
		\hline
		Parameters              &  CL     &C      &CBSL     &CBS     \\
		\hline
		{\boldmath$\Omega_b h^2   $} & $0.02228\pm 0.00016 $     & $0.02253\pm 0.00016        $    &$0.02246\pm 0.00013        $    & $0.02250\pm 0.00014        $         \\
		
		{\boldmath$\Omega_c h^2   $} & $0.1186\pm 0.0014          $    & $0.1186\pm 0.0015          $   & $0.11901\pm 0.00096        $   &  $0.11910\pm 0.00099        $                                                     \\
		
		{\boldmath$100\theta_{MC} $} & $1.04107\pm 0.00032        $   & $1.04109\pm 0.00033        $ & $1.04100\pm 0.00029        $  & $1.04102\pm 0.00030        $                                      \\
		
		{\boldmath$\tau           $} & $0.0495^{+0.0083}_{-0.0075}$     &$0.0517\pm 0.0077          $  & $0.0494^{+0.0082}_{-0.0074}$   & $0.0511\pm 0.0077          $                                   \\
		
		{\boldmath${\rm{ln}}(10^{10} A_s)$}  & $3.030^{+0.018}_{-0.015}   $  & $3.036\pm 0.016            $   &$3.031\pm 0.017            $   & $3.036\pm 0.016            $                                      \\
		
		{\boldmath$n_s            $} & $0.9684\pm 0.0046          $   & $0.9689\pm 0.0047          $        & $0.9671\pm 0.0037          $  &$0.9677\pm 0.0039          $                                        \\
		
		{\boldmath$\gamma$}     & $0.506\pm 0.022            $    & $0.468^{+0.017}_{-0.029}   $   & $0.511^{+0.022}_{-0.019}   $  & $0.470^{+0.017}_{-0.029}   $                                    \\

		\hline
		
		{\boldmath$H_0                       $ }& $68.02\pm 0.64             $ & $68.03\pm 0.67             $   & $67.80\pm 0.43             $    & $67.82\pm 0.44             $                       \\
		
		{\boldmath$\Omega_m                  $ }& $0.3064\pm 0.0085          $  & $0.3066\pm 0.0089          $     & $0.3092\pm 0.0057          $   & $0.3093\pm 0.0059          $                                                      \\
		
		{\boldmath$\sigma_8                  $ }& $0.8149\pm 0.0062          $  & $0.8299^{+0.0092}_{-0.0080}$         & $0.8149\pm 0.0063          $ & $0.8302^{+0.0093}_{-0.0082}$                                                    \\
		{\boldmath$S_8                  $ }& $0.823\pm 0.013            $  & $0.839\pm 0.016            $         & $0.827\pm 0.010            $ & $0.843\pm 0.012            $                                                    \\
		\hline
		\hline
	\end{tabular}
	\label{tab:gamma}}
\end{table*}
\begin{table*}[!t]
	\renewcommand\arraystretch{1.6}
	\caption{Mean values and $68\%$~CL marginalized errors of the most relevant cosmological parameters within the growth model parameterized via $\mu$ and $\eta$, see Eqs.~(\ref{eq:mueta}), obtained by using the CMB plus lensing, CMB alone, CMB plus BAO plus Supernovae Ia plus lensing and CMB plus BAO plus Supernovae Ia datasets, respectively.}
	\setlength{\tabcolsep}{6mm}{
		\begin{tabular} { l |c| c |c| c }
			\hline
			\hline
			Parameters              &  CL     &C      &CBSL     &CBS     \\
			\hline
			{\boldmath$\Omega_b h^2   $} & $0.02251\pm 0.00016        $     & $0.02257\pm 0.00017        $    &$0.02247\pm 0.00013        $    & $0.02254\pm 0.00014        $         \\
			
			{\boldmath$\Omega_c h^2   $} & $0.1184\pm 0.0014          $    & $0.1182\pm 0.0015          $   & $0.11885\pm 0.00094        $   &  $0.1188\pm 0.0010          $                                                    \\
			
			{\boldmath$100\theta_{MC} $} & $1.04109\pm 0.00032        $   & $1.04112\pm 0.00033        $ & $1.04102\pm 0.00029        $  & $1.04103^{+0.00026}_{-0.00031}$                                      \\
			
			{\boldmath$\tau           $} & $0.0491^{+0.0082}_{-0.0074}$     &$0.0508\pm 0.0082          $  & $0.0484^{+0.0083}_{-0.0073}$   & $0.0497^{+0.0084}_{-0.0071}$                                   \\
			
			{\boldmath${\rm{ln}}(10^{10} A_s)$}  & $3.029\pm 0.017            $  & $3.033\pm 0.017            $   &$3.028^{+0.017}_{-0.015}   $   & $3.033^{+0.018}_{-0.015}   $                                     \\
			
			{\boldmath$n_s            $} & $0.9691\pm 0.0045          $   & $0.9701\pm 0.0050          $        & $0.9679\pm 0.0037          $  &$0.9684\pm 0.0039          $                                        \\

			{\boldmath$\mu_0-1$}     & $0.09^{+0.28}_{-0.40}$    & $0.13^{+0.28}_{-0.52}   $   & $0.06^{+0.27}_{-0.38}$  & $0.08^{+0.27}_{-0.47}$                                    \\

			{\boldmath$\eta_0-1$}     & $0.23^{+0.57}_{-0.91}$    & $0.56^{+0.70}_{-1.30}   $   & $0.26^{+0.57}_{-0.92}$  & $0.62^{+0.84}_{-1.10}$                                    \\			
			
			\hline
			
			{\boldmath$H_0                       $ }& $68.11\pm 0.64             $ & $68.21\pm 0.71             $   & $67.88\pm 0.42             $    & $67.89\pm 0.45             $                       \\
			
			{\boldmath$\Omega_m                  $ }& $0.3052\pm 0.0084          $  & $0.3042\pm 0.0094          $    & $0.3082\pm 0.0056          $   & $0.3084\pm 0.0060 $                                                      \\
			
			{\boldmath$\sigma_8                  $ }& $0.810^{+0.032}_{-0.038}   $  & $0.815^{+0.031}_{-0.053}   $         & $0.808^{+0.030}_{-0.038}   $ & $0.813^{+0.029}_{-0.049}   $                                                    \\
			
			{\boldmath$S_8                  $ }& $0.817\pm 0.035            $  & $0.821^{+0.039}_{-0.051}   $         & $0.819\pm 0.034            $ & $0.824^{+0.033}_{-0.050}   $                                                   \\
			\hline
			\hline
		\end{tabular}
		\label{tab:mueta}}
\end{table*}

\begin{table*}[!t]
	\renewcommand\arraystretch{1.6}
	\caption{Mean values and $68\%$~CL marginalized errors of the most relevant cosmological parameters within the growth model parameterized via $\gamma$ and $\eta$, see Eq.~(\ref{eq:gamma}) and the second of Eqs.~(\ref{eq:mueta}), obtained by using the CMB plus lensing, CMB alone, CMB plus BAO plus Supernovae Ia plus lensing and CMB plus BAO plus Supernovae Ia datasets, respectively.}
	\setlength{\tabcolsep}{6mm}{
		\begin{tabular} { l |c| c |c| c }
			\hline
			\hline
			Parameters              &  CL     &C      &CBSL     &CBS     \\
			\hline
			{\boldmath$\Omega_b h^2   $} & $0.02250\pm 0.00016        $     & $0.02256\pm 0.00017        $    &$0.02246\pm 0.00014        $    & $0.02252\pm 0.00014        $         \\
			
			{\boldmath$\Omega_c h^2   $} & $0.1185\pm 0.0014          $    & $0.1183\pm 0.0015          $   & $0.11901\pm 0.00098        $   &  $0.1190\pm 0.0010          $                                                    \\
			
			{\boldmath$100\theta_{MC} $} & $1.04107\pm 0.00032        $   & $1.04111\pm 0.00033        $ & $1.04101\pm 0.00029        $  & $1.04106\pm 0.00029        $                                      \\
			
			{\boldmath$\tau           $} & $0.0489\pm 0.0084          $     &$0.0501\pm 0.0083          $  & $0.0486^{+0.0082}_{-0.0073}$   & $0.0505\pm 0.0079          $                                   \\
			
			{\boldmath${\rm{ln}}(10^{10} A_s)$}  & $3.029\pm 0.017            $  & $3.032\pm 0.018            $   &$3.029\pm 0.017            $   & $3.034\pm 0.017            $                                     \\
			
			{\boldmath$n_s            $} & $0.9687\pm 0.0046          $   & $0.9696\pm 0.0051          $        & $0.9674\pm 0.0038          $  &$0.9683\pm 0.0038          $                                        \\

			{\boldmath$\gamma$}     & $0.513^{+0.130}_{-0.088}    $    & $0.532^{+0.130}_{-0.065}    $   & $0.513^{+0.120}_{-0.086}    $  & $0.529^{+0.130}_{-0.071}    $                                    \\
			
			{\boldmath$\eta_0-1$}     & $0.14^{+0.47}_{-1.00}$    & $0.64^{+0.97}_{-0.75}   $   & $0.10^{+0.46}_{-0.94}$  & $0.57^{+0.88}_{-0.79}$                                    \\			
			
			\hline
			
			{\boldmath$H_0                       $ }& $68.06\pm 0.65             $ & $68.17\pm 0.70             $   & $67.81\pm 0.44             $    & $67.88\pm 0.45             $                       \\
			
			{\boldmath$\Omega_m                  $ }& $0.3058\pm 0.0086          $  & $0.3047\pm 0.0093          $    & $0.3091\pm 0.0059          $   & $0.3085\pm 0.0060$                                                      \\
			
			{\boldmath$\sigma_8                  $ }& $0.813^{+0.030}_{-0.040}   $  & $0.808^{+0.021}_{-0.042}   $         & $0.815^{+0.028}_{-0.038}   $ & $0.812^{+0.022}_{-0.042}   $                                                    \\
			
			{\boldmath$S_8                  $ }& $0.821\pm 0.037            $  & $0.815^{+0.029}_{-0.047}   $         & $0.827^{+0.032}_{-0.038}   $ & $0.823^{+0.026}_{-0.044}   $                                                   \\
			\hline
			\hline
		\end{tabular}
		\label{tab:gammaeta}}
\end{table*}

\begin{table*}[!t]
	\renewcommand\arraystretch{1.6}
	\caption{Mean values and $68\%$~CL marginalized errors of the most relevant cosmological parameters within the $\Lambda$CDM cosmology when the lensing amplitude is also a freely varying parameter. We report the results obtained by using the CMB plus lensing, CMB alone, CMB plus BAO plus Supernovae Ia plus lensing and CMB plus BAO plus Supernovae Ia datasets, respectively.}
	\setlength{\tabcolsep}{6mm}{
		\begin{tabular} { l |c| c |c| c }
			\hline
			\hline
			Parameters              &  CL     &C      &CBSL     &CBS     \\
			\hline
			{\boldmath$\Omega_b h^2   $} & $0.02251\pm 0.00017        $     & $0.02259\pm 0.00017        $    &$0.02247\pm 0.00014        $    & $0.02254\pm 0.00014        $         \\
			
			{\boldmath$\Omega_c h^2   $} & $0.1182\pm 0.0015          $    & $0.1181\pm 0.0016          $   & $0.11894\pm 0.00098        $   &  $0.11887\pm 0.00095        $                                                     \\
			
			{\boldmath$100\theta_{MC} $} & $1.04110\pm 0.00032        $   & $1.04114\pm 0.00032        $ & $1.04102\pm 0.00029        $  & $1.04105\pm 0.00029        $                                      \\
			
			{\boldmath$\tau           $} & $0.0491^{+0.0086}_{-0.0076}$     &$0.0492^{+0.0088}_{-0.0073}$  & $0.0493^{+0.0087}_{-0.0073}$   & $0.0501\pm 0.0086          $                                   \\
			
			{\boldmath${\rm{ln}}(10^{10} A_s)$}  & $3.029^{+0.018}_{-0.016}   $  & $3.029^{+0.018}_{-0.015}   $   &$3.030^{+0.018}_{-0.015}   $   & $3.033\pm 0.018            $                                      \\
			
			{\boldmath$n_s            $} & $0.9696\pm 0.0048          $   & $0.9708\pm y0.0048          $        & $0.9679\pm 0.0038          $  &$0.9688\pm 0.0037          $                                        \\
			
			{\boldmath$A_L$}     & $1.071^{+0.038}_{-0.042}   $    & $1.180\pm 0.065            $   & $1.060^{+0.032}_{-0.037}   $  & $1.158^{+0.056}_{-0.063}   $                                    \\

			\hline
			
			{\boldmath$H_0                       $ }& $68.16\pm 0.70             $ & $68.28\pm 0.72             $   & $67.85\pm 0.46             $    & $67.94\pm 0.45             $                       \\
			
			{\boldmath$\Omega_m                  $ }& $0.3045\pm 0.0092          $  & $0.3033\pm 0.0094          $     & $0.3087\pm 0.0060          $   & $0.3078\pm 0.0058$                                                      \\
			
			{\boldmath$\sigma_8                  $ }& $0.7999\pm 0.0086          $  & $0.7997\pm 0.0090          $         & $0.8026\pm 0.0078          $ & $0.8034\pm 0.0077          $                                                    \\
			{\boldmath$S_8                  $ }& $0.806\pm 0.019            $  & $0.804\pm 0.019            $         & $0.814\pm 0.013            $ & $0.814\pm 0.013            $                                                    \\
			\hline
			\hline
		\end{tabular}
		\label{tab:lcdmal}}
\end{table*}

\begin{table*}[!t]
	\renewcommand\arraystretch{1.6}
	\caption{Mean values and $68\%$~CL marginalized errors of the most relevant cosmological parameters within the growth model parameterized via one single parameter $\gamma$, when the lensing amplitude is also a freely varying parameter. We report the results obtained by using the CMB plus lensing, CMB alone, CMB plus BAO plus Supernovae Ia plus lensing and CMB plus BAO plus Supernovae Ia datasets, respectively.}
	\setlength{\tabcolsep}{6mm}{
		\begin{tabular} { l |c| c |c| c }
			\hline
			\hline
			Parameters              &  CL     &C      &CBSL     &CBS     \\
			\hline
			{\boldmath$\Omega_b h^2   $} & $0.02249\pm 0.00017        $     & $0.02258\pm 0.00017        $    &$0.02246^{+0.00013}_{-0.00015}$    & $0.02254\pm 0.00014        $         \\
			
			{\boldmath$\Omega_c h^2   $} & $0.1185\pm 0.0015          $    & $0.1182\pm 0.0016          $   & $0.11907\pm 0.00097        $   &  $0.1188\pm 0.0010          $                                                    \\
			
			{\boldmath$100\theta_{MC} $} & $1.04106\pm 0.00032        $   & $1.04114\pm 0.00033        $ & $1.04100\pm 0.00030        $  & $1.04104\pm 0.00031        $                                      \\
			
			{\boldmath$\tau           $} & $0.0506^{+0.0078}_{-0.0071}$     &$0.0525^{+0.0088}_{-0.0067}$  & $0.0495^{+0.0088}_{-0.0077}$   & $0.0490^{+0.0083}_{-0.0074}$                                   \\
			
			{\boldmath${\rm{ln}}(10^{10} A_s)$}  & $3.032\pm 0.017            $  & $3.037^{+0.019}_{-0.014}   $   &$3.032^{+0.018}_{-0.016}   $   & $3.031^{+0.018}_{-0.016}   $                                      \\
			
			{\boldmath$n_s            $} & $0.9687\pm 0.0048          $   & $0.9701\pm 0.0050          $        & $0.9672\pm 0.0039          $  &$0.9685\pm 0.0038          $                                        \\
			
			{\boldmath$A_L$}     & $0.991^{+0.057}_{-0.075}   $    & $1.085^{+0.075}_{-0.098}   $   & $0.984^{+0.056}_{-0.071}   $  & $1.086^{+0.080}_{-0.097}   $                                    \\	
					
			{\boldmath$\gamma$}     & $0.501\pm 0.036            $    & $0.497^{+0.036}_{-0.045}   $   & $0.511^{+0.022}_{-0.019}   $  & $0.503^{+0.041}_{-0.047}   $                                    \\

			\hline
			
			{\boldmath$H_0                       $ }& $68.03\pm 0.69             $ & $68.23\pm 0.72             $   & $67.79\pm 0.45             $    & $67.94\pm 0.47             $                       \\
			
			{\boldmath$\Omega_m                  $ }& $0.3062\pm 0.0092          $  & $0.3041\pm 0.0095          $    & $0.3095\pm 0.0059          $   & $0.3077\pm 0.0061$                                                      \\
			
			{\boldmath$\sigma_8                  $ }& $0.817^{+0.016}_{-0.014}   $  & $0.820^{+0.017}_{-0.013}   $         & $0.819^{+0.015}_{-0.013}   $ & $0.817\pm 0.015            $                                                    \\
			
			{\boldmath$S_8                  $ }& $0.826\pm 0.022            $  & $0.825^{+0.024}_{-0.022}   $         & $0.832\pm 0.018            $ & $0.827\pm 0.019            $                                                   \\
			\hline
			\hline
		\end{tabular}
		\label{tab:gammaal}}
\end{table*}

\begin{table*}[!t]
	\renewcommand\arraystretch{1.6}
	\caption{Mean values and $68\%$~CL marginalized errors of the most relevant cosmological parameters within the growth model parameterized via $\mu$ and $\eta$, see Eq.~(\ref{eq:mueta}), when the lensing amplitude is also a freely varying parameter. We report the results obtained by using the CMB plus lensing, CMB alone, CMB plus BAO plus Supernovae Ia plus lensing and CMB plus BAO plus Supernovae Ia datasets, respectively.}
	\setlength{\tabcolsep}{6mm}{
		\begin{tabular} { l |c| c |c| c }
			\hline
			\hline
			Parameters              &  CL     &C      &CBSL     &CBS     \\
			\hline
			{\boldmath$\Omega_b h^2   $} & $0.02249\pm 0.00017        $     & $0.02259\pm 0.00017        $    &$0.02246\pm 0.00014        $    & $0.02252\pm 0.00014        $         \\
			
			{\boldmath$\Omega_c h^2   $} & $0.1186\pm 0.0015          $    & $0.1181\pm 0.0015          $   & $0.1190\pm 0.0010          $   &  $0.11884\pm 0.00099        $                                                    \\
			
			{\boldmath$100\theta_{MC} $} & $1.04107\pm 0.00033        $   & $1.04115\pm 0.00032        $ & $1.04100\pm 0.00029        $  & $1.04104\pm 0.00030        $                                      \\
			
			{\boldmath$\tau           $} & $0.0496^{+0.0084}_{-0.0073}$     &$0.0498^{+0.0085}_{-0.0075}$  & $0.0493\pm 0.0085          $   & $0.0494\pm 0.0082          $                                  \\
			
			{\boldmath${\rm{ln}}(10^{10} A_s)$}  & $3.030^{+0.018}_{-0.015}   $  & $3.030\pm 0.017            $   &$3.031^{+0.018}_{-0.016}   $   & $3.032\pm 0.018            $                                     \\
			
			{\boldmath$n_s            $} & $0.9685\pm 0.0049          $   & $0.9706\pm 0.0048          $        & $0.9675\pm 0.0039          $  &$0.9688\pm 0.0039          $                                        \\
			
			{\boldmath$A_L$}     & $0.958^{+0.062}_{-0.080}   $    & $1.062^{+0.082}_{-0.110}    $   & $0.08^{+0.25}_{-0.48}$  & $1.041^{+0.082}_{-0.100}    $                                      \\
			
			{\boldmath$\mu_0-1$}     & $0.12^{+0.29}_{-0.44}$    & $0.11^{+0.28}_{-0.46}$   & $0.06^{+0.27}_{-0.38}$  & $0.12^{+0.29}_{-0.46}$                                    \\
			
			{\boldmath$\eta_0-1$}     & $0.35^{+0.54}_{-1.10}$    & $0.39^{+0.59}_{-1.10}   $   & $0.47^{+0.60}_{-1.20}$  & $0.40^{+0.59}_{-1.20}$                                    \\			
			
			\hline
			
			{\boldmath$H_0                       $ }& $68.00\pm 0.69             $ & $68.29\pm 0.71             $   & $67.80\pm 0.46             $    & $67.94\pm 0.46             $                       \\
			
			{\boldmath$\Omega_m                  $ }& $0.3066\pm 0.0092          $  & $0.3031\pm 0.0093          $    & $0.3092\pm 0.0061          $   & $0.3078\pm 0.0060$                                                      \\
			
			{\boldmath$\sigma_8                  $ }& $0.814^{+0.032}_{-0.045}   $  & $0.812^{+0.031}_{-0.046}   $         & $0.812^{+0.028}_{-0.047}   $ & $0.815^{+0.031}_{-0.047}   $                                                    \\
			
			{\boldmath$S_8                  $ }& $0.823^{+0.038}_{-0.046}   $  & $0.816^{+0.037}_{-0.046}   $         & $0.824^{+0.032}_{-0.048}   $ & $0.826^{+0.034}_{-0.047}   $                                                   \\
			\hline
			\hline
		\end{tabular}
		\label{tab:muetaal}}
\end{table*}

\begin{table*}[!t]
	\renewcommand\arraystretch{1.6}
	\caption{Mean values and $68\%$~CL marginalized errors of the most relevant cosmological parameters within the growth model parameterized via $\gamma$ and $\eta$, see Eqs.~(\ref{eq:gamma}) and the second of Eqs.~(\ref{eq:mueta}), when the lensing amplitude is also a freely varying parameter. We report the results obtained by using the CMB plus lensing, CMB alone, CMB plus BAO plus Supernovae Ia plus lensing and CMB plus BAO plus Supernovae Ia datasets, respectively.}
	\setlength{\tabcolsep}{6mm}{
		\begin{tabular} { l |c| c |c| c }
			\hline
			\hline
			Parameters              &  CL     &C      &CBSL     &CBS     \\
			\hline
			{\boldmath$\Omega_b h^2   $} & $0.02248\pm 0.00016        $     & $0.02260\pm 0.00017        $    &$0.02245\pm 0.00014        $    & $0.02254\pm 0.00014        $         \\
			
			{\boldmath$\Omega_c h^2   $} & $0.1186\pm 0.0015          $    & $0.1180\pm 0.0016          $   & $0.1191\pm 0.0010          $   &  $0.1189\pm 0.0010          $                                                    \\
			
			{\boldmath$100\theta_{MC} $} & $1.04107\pm 0.00032        $   & $1.04115\pm 0.00033        $ & $1.04101\pm 0.00029        $  & $1.04104\pm 0.00029        $                                      \\
			
			{\boldmath$\tau           $} & $0.0492^{+0.0086}_{-0.0074}$     &$0.0494^{+0.0086}_{-0.0077}$  & $0.0492^{+0.0083}_{-0.0074}$   & $0.0500\pm 0.0078          $                                   \\
			
			{\boldmath${\rm{ln}}(10^{10} A_s)$}  & $3.029^{+0.018}_{-0.016}   $  & $3.030^{+0.018}_{-0.016}   $   &$3.031\pm 0.017            $   & $3.033\pm 0.017            $                                     \\
			
			{\boldmath$n_s            $} & $0.9684\pm 0.0047          $   & $0.9706\pm 0.0049          $        & $0.9674\pm 0.0038          $  &$0.9687\pm 0.0039          $                                       \\
			
			{\boldmath$A_L$}     & $0.971^{+0.063}_{-0.078}   $    & $1.078^{+0.080}_{-0.110}    $   & $0.962^{+0.060}_{-0.073}   $  & $1.059^{+0.080}_{-0.093}   $                                    \\
			
			{\boldmath$\gamma$}     & $0.527^{+0.130}_{-0.071}    $    & $0.528^{+0.120}_{-0.076}    $   & $0.523^{+0.130}_{-0.075}    $  & $0.526^{+0.110}_{-0.077}    $                                    \\
			
			{\boldmath$\eta_0-1$}     & $0.36^{+0.65}_{-1.00}$    & $0.36^{+0.63}_{-1.00}$   & $0.35^{+0.60}_{-1.10}$  & $0.36^{+0.71}_{-0.91}$                                    \\			
			
			\hline
			
			{\boldmath$H_0                       $ }& $67.99\pm 0.68             $ & $68.32\pm 0.72             $   & $67.79\pm 0.45             $    & $67.94\pm 0.45             $                       \\
			
			{\boldmath$\Omega_m                  $ }& $0.3067\pm 0.0091          $  & $0.3029\pm 0.0094          $    & $0.3095\pm 0.0060          $   & $0.3078\pm 0.0060$                                                      \\
			
			{\boldmath$\sigma_8                  $ }& $0.810^{+0.024}_{-0.041}   $  & $0.808^{+0.024}_{-0.038}   $         & $0.813^{+0.023}_{-0.041}   $ & $0.812^{+0.024}_{-0.038}   $                                                    \\
			
			{\boldmath$S_8                  $ }& $0.819^{+0.032}_{-0.044}   $  & $0.811^{+0.031}_{-0.041}   $         & $0.825^{+0.027}_{-0.042}   $ & $0.822^{+0.027}_{-0.040}   $                                                   \\
			\hline
			\hline
		\end{tabular}
		\label{tab:gammaetaal}}
\end{table*}

\begin{table*}[!t]
	\renewcommand\arraystretch{1.6}
	\caption{Mean values and $68\%$~CL marginalized errors of the most relevant cosmological parameters within the $\Lambda$CDM cosmology when the lensing amplitude is an oscillatory function of the multipole $\ell$, $A_\ell= A_{\rm L} + A_m \sin \ell$. We report the results obtained by using the CMB plus lensing, CMB alone, CMB plus BAO plus Supernovae Ia plus lensing and CMB plus BAO plus Supernovae Ia datasets, respectively.}
	\setlength{\tabcolsep}{6mm}{
		\begin{tabular} { l |c| c |c| c }
			\hline
			\hline
			Parameters              &  CL     &C      &CBSL     &CBS     \\
			\hline
			{\boldmath$\Omega_b h^2   $} & $0.02250\pm 0.00017        $     & $0.02260\pm 0.00017        $    &$0.02246\pm 0.00014        $    & $0.02253\pm 0.00014        $         \\
			
			{\boldmath$\Omega_c h^2   $} & $0.1184\pm 0.0015          $    & $0.1179\pm 0.0015          $   & $0.11892\pm 0.00099        $   &  $0.1188\pm 0.0010          $                                                    \\
			
			{\boldmath$100\theta_{MC} $} & $1.04109\pm 0.00033        $   & $1.04116\pm 0.00033        $ & $1.04103\pm 0.00029        $  & $1.04103\pm 0.00030        $                                      \\
			
			{\boldmath$\tau           $} & $0.0489^{+0.0085}_{-0.0077}$     &$0.0489\pm 0.0086          $  & $0.0490\pm 0.0079          $   & $0.0487\pm 0.0081          $                                   \\
			
			{\boldmath${\rm{ln}}(10^{10} A_s)$}  & $3.028\pm 0.018            $  & $3.028^{+0.018}_{-0.016}   $   &$3.030\pm 0.016            $   & $3.030^{+0.018}_{-0.016}   $                                     \\
			
			{\boldmath$n_s            $} & $0.9695\pm 0.0049          $   & $0.9713\pm 0.0049          $        & $0.9683\pm 0.0037          $  &$0.9689\pm 0.0038          $                                      \\
			
			{\boldmath$A_L$}     & $1.093\pm 0.045            $    & $1.21\pm 0.17              $   & $1.085\pm 0.040            $  & $1.15\pm 0.17              $                                    \\
			
			{\boldmath$A_m$}     & $0.070\pm 0.057            $    & $0.08\pm 0.47              $   & $0.071\pm 0.055            $  & $-0.04\pm 0.51             $                                    \\			
			
			\hline
			
			{\boldmath$H_0                       $ }& $68.10\pm 0.71             $ & $68.36\pm 0.71             $   & $67.85\pm 0.45             $    & $67.94\pm 0.46             $                       \\
			
			{\boldmath$\Omega_m                  $ }& $0.3054\pm 0.0093          $  & $0.3023\pm 0.0093          $    & $0.3086\pm 0.0059          $   & $0.3078\pm 0.0061$                                                      \\
			
			{\boldmath$\sigma_8                  $ }& $0.8003\pm 0.0089          $  & $0.7989^{+0.0094}_{-0.0084}$         & $0.8024\pm 0.0074          $ & $0.8023\pm 0.0078          $                                                    \\
			
			{\boldmath$S_8                  $ }& $0.807\pm 0.019            $  & $0.802\pm 0.019            $         & $0.814\pm 0.013            $ & $0.813\pm 0.014            $                                                   \\
			\hline
			\hline
		\end{tabular}
		\label{tab:lcdmosc}}
\end{table*}

\begin{table*}[!t]
	\renewcommand\arraystretch{1.6}
	\caption{Mean values and $68\%$~CL marginalized errors of the most relevant cosmological parameters within the growth model parameterized via $\mu$ and $\eta$, see Eqs.~(\ref{eq:mueta}), when the lensing amplitude is an oscillatory function of the multipole $\ell$, $A_\ell= A_{\rm L} + A_m \sin \ell$. We report the results obtained by using the CMB plus lensing, CMB alone, CMB plus BAO plus Supernovae Ia plus lensing and CMB plus BAO plus Supernovae Ia datasets, respectively.}
	\setlength{\tabcolsep}{6mm}{
		\begin{tabular} { l |c| c |c| c }
			\hline
			\hline
			Parameters              &  CL     &C      &CBSL     &CBS     \\
			\hline
			{\boldmath$\Omega_b h^2   $} & $0.02250\pm 0.00017        $     & $0.02260\pm 0.00018        $    &$0.02245\pm 0.00014        $    & $0.02253\pm 0.00014        $         \\
			
			{\boldmath$\Omega_c h^2   $} & $0.1186\pm 0.0015          $    & $0.1180\pm 0.0015          $   & $0.1191\pm 0.0010          $   &  $0.1189\pm 0.0011          $                                                    \\
			
			{\boldmath$100\theta_{MC} $} & $1.04107\pm 0.00032        $   & $1.04116\pm 0.00034        $ & $1.04102\pm 0.00029        $  & $1.04104\pm 0.00032        $                                      \\
			
			{\boldmath$\tau           $} & $0.0494^{+0.0087}_{-0.0072}$     &$0.0505^{+0.0090}_{-0.0079}$  & $0.0492\pm 0.0081          $   & $0.0496^{+0.0082}_{-0.0073}$                                   \\
			
			{\boldmath${\rm{ln}}(10^{10} A_s)$}  & $3.030^{+0.018}_{-0.016}   $  & $3.032^{+0.020}_{-0.016}   $   &$3.030\pm 0.017            $   & $3.032^{+0.018}_{-0.015}   $                                     \\
			
			{\boldmath$n_s            $} & $0.9687\pm 0.0048          $   & $0.9707\pm 0.0049          $        & $0.9676\pm 0.0038          $  &$0.9686\pm 0.0041          $                                      \\
			
			{\boldmath$A_L$}     & $0.988^{+0.077}_{-0.087}   $    & $1.07\pm 0.16              $   & $0.980^{+0.072}_{-0.080}   $  & $1.05\pm 0.18              $                                    \\
			
			{\boldmath$A_m$}     & $0.041\pm 0.056            $    & $0.01\pm 0.42              $   & $0.041\pm 0.055            $  & $0.01^{+0.63}_{-0.53}      $                                    \\			
			
			{\boldmath$\mu_0-1$}     & $0.05^{+0.24}_{-0.44}$    & $0.07^{+0.26}_{-0.39}$   & $0.08^{+0.25}_{-0.45}$  & $0.09^{+0.26}_{-0.47}$                                    \\

            {\boldmath$\eta_0-1$}     & $0.43^{+0.58}_{-1.10}$    & $0.44^{+0.62}_{-0.96}$   & $0.37^{+0.59}_{-1.10}$  & $0.42^{+0.71}_{-1.10}$                                    \\			
			
			\hline
			
			{\boldmath$H_0                       $ }& $68.00\pm 0.68             $ & $68.33\pm 0.71             $   & $67.79\pm 0.45             $    & $67.90\pm 0.49             $                       \\
			
			{\boldmath$\Omega_m                  $ }& $0.3067\pm 0.0091          $  & $0.3027\pm 0.0092          $    & $0.3094\pm 0.0061          $   & $0.3083\pm 0.0065$                                                      \\
			
			{\boldmath$\sigma_8                  $ }& $0.807^{+0.027}_{-0.044}   $  & $0.808^{+0.027}_{-0.040}   $         & $0.811^{+0.027}_{-0.045}   $ & $0.813^{+0.028}_{-0.048}   $                                                    \\
			
			{\boldmath$S_8                  $ }& $0.816^{+0.035}_{-0.045}   $  & $0.812^{+0.032}_{-0.043}   $         & $0.824^{+0.032}_{-0.046}   $ & $0.824^{+0.031}_{-0.050}   $                                                   \\
			\hline
			\hline
		\end{tabular}
		\label{tab:muetaosc}}
\end{table*}

\section{Acknowledgements}
This work has been supported by the Spanish MCIN/AEI/10.13039/501100011033 grants PID2020-113644GB-I00 and by the European ITN project HIDDeN (H2020-MSCA-ITN-2019/860881-HIDDeN) and SE project ASYMMETRY (HORIZON-MSCA-2021-SE-01/101086085-ASYMMETRY) and well as by the Generalitat Valenciana grants PROMETEO/2019/083 and CIPROM/2022/69.
DW is supported by the CDEIGENT Project of Consejo Superior de Investigaciones Científicas (CSIC).

\bibliography{growth}

\end{document}